\documentclass[superscriptaddress,preprint,nofootinbib,nobibnotes,eprint,amsmath,amssymb,aps,prb,citeautoscript,titlepage,nobalancelastpage,floats,final,eqsecnum]{revtex4-2}

\usepackage[utf8]{inputenc}

\usepackage{amsmath}
\usepackage{amssymb}
\usepackage{amsfonts}

\usepackage{latexsym}
\usepackage{slashed}
\usepackage[svgnames]{xcolor}
\usepackage{soul}
\usepackage{braket}
\usepackage{MnSymbol}
\usepackage{wasysym}
\usepackage{mathrsfs}  
\usepackage{physics}
\usepackage{enumitem}        

\usepackage{graphics}
\graphicspath{ {graphs/} }

\usepackage{microtype}
\usepackage{multirow}

\usepackage{hyperref}
\usepackage{natbib}

\hypersetup{
	pdfencoding=auto,
	psdextra,
	colorlinks=true,
	linkcolor=Blue,
	citecolor=DarkRed,
	urlcolor=Blue
}

\usepackage{mathtools}

\usepackage{easyReview}

\begin{document}

	\title{\Large{{\sc Generalized Modulated Symmetries in \\ $\mathbb{Z}_2$ Topological Ordered Phases}}}

	\author{Gustavo M. Yoshitome}
	\email{gustavo.yoshitome@uel.br }
	\affiliation{Departamento de Física, Universidade Estadual de Londrina, Londrina, PR, Brasil}
	
	\author{Heitor Casasola}
	\email{heitor.casasola@uel.br}
	\affiliation{Departamento de Física, Universidade Estadual de Londrina, Londrina, PR, Brasil}
	
	\author{Rodrigo Corso}
	\email{rodrigocorso@uel.br}
	\affiliation{Departamento de Física, Universidade Estadual de Londrina, Londrina, PR, Brasil}
	
	\author{Pedro R. S. Gomes}
	\email{pedrogomes@uel.br}
	\affiliation{Departamento de Física, Universidade Estadual de Londrina, Londrina, PR, Brasil}

	\date{\today}
	
	\begin{abstract}

		We study $\mathbb{Z}_2$ topological ordered phases in 2+1 dimensions characterized by generalized modulated symmetries. Such phases have explicit realizations in terms of fixed-point Hamiltonians involving commuting projectors with support $h=3,5,7,\ldots$ in the horizontal direction, which dictates the modulation of the generalized symmetries. These symmetries are sensitive to the lattice sizes. For certain sizes, they are spontaneously broken and the ground state is degenerated, while for the remaining ones, the symmetries are explicitly broken and the ground state is unique. The ground state dependence on the lattice sizes is a manifestation of the ultraviolet/infrared (UV/IR) mixing. The structure of the modulated symmetries implies that the anyons can move only in  rigid steps of size $h$, leading to the notion of position-dependent anyons. The phases exhibit rich boundary physics with a variety of gapped phases, including trivial, partial and total symmetry-breaking, and SPT phases. Effective field theory descriptions are discussed, making transparent the relation between the generalized modulated symmetries and the restrictions on anyon mobility, incorporating the boundary physics in a natural way, and showing how the short-distance details can be incorporated into the continuum by means of twisted boundary conditions.

	\end{abstract}

	\maketitle

	\tableofcontents

	\section{Introduction}

	Symmetry and topological order are among the most profound and comprehensive concepts of physics. Systems exhibiting an interplay between these two ingredients are often endowed with spectacular physical properties. Generalized symmetries, which are one of the central subjects of contemporary physics, originated from such interplay \cite{Nussinov:2006iva,Nussinov_2009}. In this context, topological order can be understood within the Landau paradigm of symmetry-based classification of phases, in the sense that topological order is characterized by the spontaneous breaking of generalized finite symmetries \cite{Gaiotto:2014kfa,Wen:2018zux,McGreevy_2023}.

	The more exotic the type of the generalized symmetry, the more dramatic the physical consequences. The quintessential example is the case of fracton topological order \cite{Chamon:2004lew,Haah:2011drr,Vijay:2016phm}, which is characterized by the spontaneous breaking of subsystem symmetries \cite{Qi:2020jrf,Rayhaun:2021ocs}.  A distinctive feature of subsystem symmetries is that the number of symmetries typically scales with the system size. This implies, in particular, a ground state degeneracy that diverges in the thermodynamic limit, which is a manifestation of ultraviolet/infrared (UV/IR) mixing. In the extreme case of type-II fracton physics, associated with fractal generalized symmetries, the ground state degeneracy is much more sensitive to the lattice sizes, not having in general a smooth dependence on the lattice size \cite{Haah_2013,Casasola:2023tot,Casasola2024}. This is a severe form of UV/IR mixing, which defies the very conception of such systems as phases of matter.

	Milder forms of UV/IR mixing show up in topological phases characterized by spatially modulated generalized symmetries, i.e., internal symmetries which act non-uniformly on the  subspaces they belong \cite{Sala:2021jca,Delfino:2023anb,Delfino:2023rpw,Pace:2024tgk,Seo:2024its}. A common feature associated with modulated symmetries is a restriction on the mobility of anyons. This is so due to the existence of transport operators that move single excitations only in steps larger than the unit lattice spacing, with the step-size dictated by the order $N$ of the symmetry $\mathbb{Z}_N$ of the local degrees of freedom. This leads to the notion of position-dependent anyons \cite{Bulmash_2018,Oh_2022,Pace_2022,Delfino_2023,Watanabe:2022pgk,Oh:2023bnk,Ebisu:2022nln,Ebisu:2023ayu,PhysRevB.106.155150,Ebisu:2023idd,PhysRevB.109.235127}.

	In this work, we study topological ordered phases in the two-dimensional square lattice exhibiting spatially modulated generalized symmetries along one of the directions. The form of these modulated symmetries is dictated by certain linear recurrence relations. Specifically, there are symmetry operators 
	\begin{equation}
		\prod_{i_x=1}^{L_x}X_{(i_x,i_y)}^{t_{(i_x,i_y)}}, 
	\end{equation}
	where $X$ is a $\mathbb{Z}_2$ Pauli operators, with the weights $t_{(i_x,i_y)}=0,1 \mod 2$ satisfying the recurrence relation
	\begin{equation}
		t_{\vec{r}}+t_{\vec{r}+\hat{x}}+ \cdots +t_{\vec{r}+(h-1)\hat{x}} = 0 \mod 2.
	\end{equation}
	The parameter $h$ corresponds to the support of the commuting projectors, 
	\begin{equation}
		P^{(h)}_{\vec{r}} = \underbrace{ \begin{matrix}
				{}&{}&{} & X &{} &{}&{}
				\\
				\cdots&Z & Z & Z_{\vec{r}} & Z &Z& \cdots
				\\
				{}&{}&{} & X &{} &{} &{}
		\end{matrix}}_{h\, \text{consecutive sites}}.
		\label{cp00}
	\end{equation}
	As the commuting projector is symmetric in relation to its central site $\vec{r}$, $h$ is odd. 
	The Hamiltonian constructed in terms of such stabilizers can also be put in the form of a CSS code through the application of Hadamard operators at specific horizontal lines (even or odd), offering a perspective from which some properties can be studied in a systematic way, as described, for example, in Ref. \cite{rmy6-9n89}.

	The resulting phases are symmetry-enriched topological (SET) phases which realize the symmetry pattern
	\begin{equation}
		\underbrace{\mathbb{Z}_2 \times \mathbb{Z}_2 \times \cdots \times\mathbb{Z}_2}_{h-1}.	
	\end{equation}
	They exhibit a mild form of UV/IR mixing, in the sense that the ground state is sensitive to the lattice sizes $L_x$ and $L_y$, but does not scale with them. Notably, for certain sizes the ground state is unique. This feature has also been observed in other topologically ordered phases \cite{Casasola:2023tot,Casasola2024,Watanabe:2022pgk}.

	In common with references \cite{Oh_2022,Pace_2022,Delfino_2023,Watanabe:2022pgk,Oh:2023bnk}, the phases discussed here are also endowed with position-dependent anyons, but with the difference that the local degrees of freedom are $\mathbb{Z}_2$, instead of $\mathbb{Z}_N$. The step-size of the transport operators is dictated by the support $h$ of the commuting projectors \eqref{cp00} entering the Hamiltonian. A topologically ordered code with the commuting projectors involving more distant sites than nearest neighbors and featuring  some similar properties has been recently studied in \cite{Chen:2025hzm}.

	
	\subsection{Outline}
	
	In the core of the manuscript, we carry out an extensive study of the simplest case $h=3$, with symmetry pattern $\mathbb{Z}_2\times \mathbb{Z}_2$, which already captures the main features of the phases\footnote{Some aspects discussed here may be facilitated with the availability of powerful computational tools, for example, as in Ref. \cite{Liang:2023lct}.}.  
	
	We begin by computing the ground state degeneracy in Sec. \ref{tripartite}, which turns out to be sensitive to the lattice sizes $L_x$ and $L_y$, signaling a UV/IR mixing. However, the ground state degeneracy does not scale with $L_x$ and $L_y$, i.e., it is finite in the thermodynamic limit. When $L_x$ is not a multiple of three, the ground state is unique, which is a peculiar feature in the context of topologically ordered phases. 
	
	In Sec. \ref{symmetryperators}, we discuss the symmetry operators and analyze the algebra between them. This allows us to understand the ground state degeneracy in terms of 't Hooft anomalies \cite{tHooft:1979rat,Kapustin:2014lwa,Kapustin:2014zva}.  We also discuss the restrictions on the mobility of the excitations.  
	
	In Sec. \ref{li00}, we discuss the topologically ordering from the perspective of spontaneous breaking of generalized modulated symmetries and compute the entanglement entropy, which has a nontrivial topological contribution. 
	
	A topologically ordered phase with the symmetry pattern $\mathbb{Z}_2\times \mathbb{Z}_2$ encodes a rich boundary phase structure, realizing the so-called topological holography \cite{Levin_2013,Lichtman:2020nuw,Moradi:2022lqp,Huang:2023pyk}. In Sec. \ref{boundarytheory0}, we construct explicitly boundary Hamiltonians corresponding to all possible condensations of anyons and characterize in great detail the boundary phases according to their symmetries. 
	A detailed algorithm for constructing gapped boundaries via Lagrangian subgroup condensation can be found in the recent study \cite{Liang:2024bez}.

	In phases exhibiting UV/IR mixing such as the ones considered here, a natural question is concerning the continuum limit. To explore this, we investigate effective field theory descriptions in two regimes. The first one, in Sec. \ref{eft01}, corresponds to a low-energy and short-distance description, in which the lattice is preserved in the discussion. This approach is enlightening in that it revels in a transparent way the role of the modulated symmetries in restricting the mobility of excitations. In addition, it incorporates the boundary physics in a very natural way. 
	
	Then, in Sec. \ref{eft02}, we consider a description in low energies and long distances. The position-dependent anyons are translated into distinct anyon species in the continuum, parameterized by a set of mutual Chern-Simons fields coupled through a $K$-matrix, which is chosen to reproduce the mutual statistics of the excitations \cite{PhysRevB.78.155134,PhysRevB.86.125101,Han:2024nvu,PhysRevB.106.155150}. In addition, the fields are supplemented with twisted boundary conditions, encoding the UV information that the superselection sectors are affected by the lattice sizes. Taking these aspects into account allows us to recover the ground state degeneracy observed in the lattice model.
	
	In Sec. \ref{hpartitecase}, we discuss briefly some properties of the topological phases with generic $h$. We compute the ground state degeneracy, discuss the symmetry operators, the restrictions on mobility of excitations, and the 't Hooft anomalies. 
	
	We conclude in Sec. \ref{finalremarks} with a brief summary and some final comments. Additional details of the boundary phases are discussed in appendix \ref{appendixA}.

	
	\section{Tripartite  Generalized Modulated Symmetries}\label{tripartite}
	
\begin{figure}
		\centering
		\includegraphics[angle=90,scale=.75]{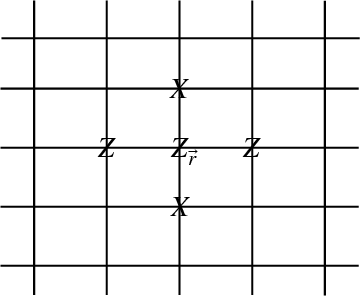}
		\caption{Definition of the commuting projector $P_{\vec{r}}$. }
		\label{fig:stabilizer}
	\end{figure}

	We consider here the simplest realization ($h=3$) of the class of phases we study in this paper. It consists of a two-dimensional model defined on the square lattice, with a  qubit attached to each site $\vec{r}= i_x \hat{x}+ i_y \hat{y} \equiv (i_x,i_y)$, governed by the fixed-point Hamiltonian
	\begin{equation}
		H = - \sum_{\vec{r}} P_{\vec{r}},
		\label{eq:model3}
	\end{equation}
	where the commuting projectors $P_{\vec{r}} \equiv 	P_{\vec{r}}^{(3)}$ are given by 
	\begin{equation}
		P_{\vec{r}} = X_{\vec{r}-\hat{y}} \, Z_{\vec{r}-\hat{x}} \, Z_{\vec{r}} \, Z_{\vec{r}+ \hat{x}}  \, X_{\vec{r}+\hat{y}}~=
		\begin{matrix}
			{}&{} & X &{} &{}
			\\
			& Z& Z_{\vec{r}} & Z &
			\\
			{}&{} & X &{} &{} 
		\end{matrix}.
	\end{equation}
	Fig. \ref{fig:stabilizer} shows this commuting projector on the lattice.
	
	The Hamiltonian \eqref{eq:model3} is exactly soluble since all the operators $P_{\vec{r}}$ are simultaneously commuting, i.e.,  $[P_{\vec{r}},P_{\vec{r}'}]=0$ for all $\vec{r}$ and $\vec{r}'$. Since $P_{\vec{r}} ^{2} = \openone$, its eigenvalues are $\pm 1$ and a ground state corresponds to a state in which
	\begin{equation}
		P_{\vec{r}} \ket{GS} =  \ket{GS}, ~~~ \text{for all} ~ \vec{r}. \label{eq:gscondition}
	\end{equation}
	
	In a lattice with periodic boundary conditions (PBC), there is a one-to-one correspondence between commuting projectors and sites. A lattice with sizes $L_x$ and  $L_y$ has a total of $2^{L_x L_y}$ states. It also has this number of eigenvalues of commuting projectors available to label such states. However, it may happen that there are global contraints between the commuting projectors so that not all eigenvalues are independent. In this case, this will imply a nontrivial ground state degeneracy, 
	\begin{equation}
		\text{GSD} = \frac{2^{L_x L_y}}{2^{L_x L_y - N_c}} = 2^{N_c},
	\end{equation}
	where $N_c$ is the total number of global constraints. If there are no constraints, the ground state is unique.
	
	To find the number of constraints, we consider their generic form 
	\begin{equation}
		\prod_{\vec{r}} ({P}_{\vec{r}})^{t_{\vec{r}}} = \openone,
		\label{generalconstraint}
	\end{equation}
	where $t_{\vec{r}}=0,1 \mod 2$. Nontrivial solutions for the set $\{ t_{\vec{r}} \}$ correspond to constraints. This relation can be rewritten as
	\begin{equation}
		\prod_{\vec{r}}  (X_{\vec{r}})^{t_{\vec{r}-\hat{y}} + t_{\vec{r}+\hat{y}}} (Z_{\vec{r}})^{t_{\vec{r}-\hat{x}} + t_{\vec{r}}+ t_{\vec{r}+\hat{x}}} = \openone,
	\end{equation}
	from which we see that 
	\begin{equation}
		t_{\vec{r}-\hat{y}} + t_{\vec{r}+\hat{y}} = 0 \mod 2 \label{eq:const1}
	\end{equation}
	and 
	\begin{equation}
		t_{\vec{r}-\hat{x}} + t_{\vec{r}}+ t_{\vec{r}+\hat{x}} = 0 \mod 2.  \label{eq:const2}
	\end{equation}
	
	It is convenient to write Eq.~\eqref{eq:const1} as
	\begin{equation}
		t_{\vec{r}} + t_{\vec{r}+2\hat{y}} = 0 \mod 2,
		\label{123}
	\end{equation}
	to make evident the structure of the constraints between $t_{\vec{r}}$'s along the $y$ direction. There are two different situations depending whether $L_y$ is even or odd. For $L_y$ even, the $t_{\vec{r}}$'s at even sites are independent of the ones at odd sites, splitting the set $\{t_{\vec{r}}\}$  into two independent sets, $\left\lbrace t_{(i_x,1)}, t_{(i_x,3)}, t_{(i_x,5)}, \dots , t_{(i_x, L_y-3)}, t_{(i_x,L_y-1)} \right\rbrace$ and $\left\lbrace t_{(i_x,2)}, t_{(i_x,4)}, t_{(i_x,6)}, \dots , t_{(i_x, L_y-4)}, t_{(i_x, L_y-2)} \right\rbrace$. In the case of $L_y$ odd, all the $t_{\vec{r}}$'s are constrained.

	\begin{figure}
		\includegraphics[width = 0.5\linewidth]{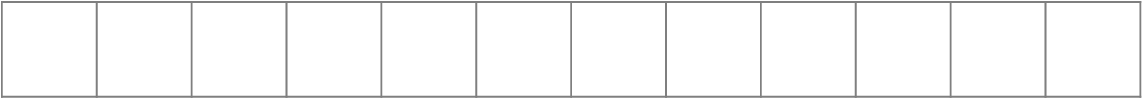}
		
		\vspace{.3cm}
		
		\includegraphics[width = 0.5 \linewidth]{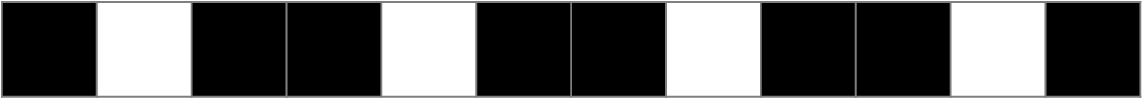}
		
		\vspace{.3cm}
		
		\includegraphics[width = 0.5 \linewidth]{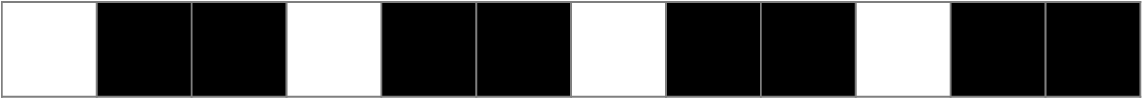}
		
		\vspace{.3cm}
		
		\includegraphics[width = 0.5 \linewidth]{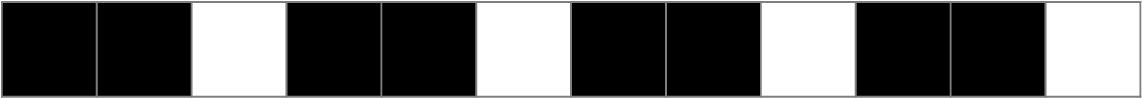}
		\caption{Representation of the solutions of the rule \eqref{eq:const2}, for $L=12$.} \label{fig:ca3solutions}
	\end{figure}
	Eq. \eqref{eq:const2} corresponds to a linear recurrence relation where the value of $t_{\vec{r}}$ in a given site depends on the values of $t_{\vec{r}}$ in the two previous sites along the $x$-direction. In Fig.~\ref{fig:ca3solutions}, we show four different solutions for \eqref{eq:const2}: the trivial solution, $t_{\vec{r}} = 0$ for every $\vec{r}$, and three nontrivial ones. However, note that only two of them are linearly independent. Picking any two solutions, the third one is obtained from their linear combination. These four different cases are spanned by the different combinations of the values of the first two sites in the $x$-direction, since $t_{\vec{r}}$  has a three-site periodicity. This follows easily by combining Eq.~\eqref{eq:const2} for different $\vec{r}$,
	\begin{equation}
		\left(	t_{\vec{r}} + t_{\vec{r}+\hat{x}}+ t_{\vec{r}+2\hat{x}}\right)  +\left( t_{\vec{r}+\hat{x}} + t_{\vec{r}+2\hat{x}}+ t_{i\vec{r}+3\hat{x}} \right) = t_{\vec{r}} + t_{\vec{r}+3\hat{x}} = 0 \mod 2.
	\end{equation}
	With this, we can write in general
	\begin{equation}
		t_{\vec{r}} = t_{\vec{r}+3 m \hat{x}} \label{eq:const3}, 
	\end{equation}
	for any integer $m$.

	In the case of $L_x$ being a multiple of three, Eq.~\eqref{eq:const3} divides the set of $t_{\vec{r}}$'s into three parts,
	\begin{align}
		&t_{(1,i_y)} = t_{(4,i_y)}  = t_{(7,i_y)}  = \, \dots \, = t_{(L_x-2,i_y)},\\
		&t_{(2,i_y)}= t_{(5,i_y)}=t_{(8,i_y)}= \, \dots \, = t_{(L_x-1,i_y)},\\
		&t_{(3,i_y)}= t_{(6,i_y)}= t_{(9,i_y)} =\, \dots \, = t_{(L_x,i_y)}.
	\end{align}
	As $t_{(3,i_y)}$ can be determined for given $t_{(1,i_y)}$ and $t_{(2,i_y)}$, there are only four different possibilities for the system, as depicted in Fig.~\ref{fig:ca3solutions}. The number of nontrivial linearly independent solutions for $t_{\vec{r}}$'s is two if $L_x$ is a multiple of three. Otherwise, there is only the trivial solution. This result can be conveniently expressed as
	\begin{equation}
		\gcd {\left(3, L_x\right)} - 1.
	\end{equation} 
	
	Combining the different possibilities for both $L_x$ and $L_y$, we obtain that the total number of constraints is
	\begin{equation}
		N_c = \gcd{\left( 2, L_y\right)} \, \left[\gcd{\left( 3, L_x \right)} - 1 \right].
	\end{equation}
	Consequently, the ground state degeneracy is
	\begin{equation}
		\text{GSD} = 2^{\gcd{\left( 2, L_y\right)} \, \left[\gcd{\left( 3, L_x \right)} - 1 \right]}.
		\label{GSD}
	\end{equation}
	Some comments are in order. First, when $L_x$ is not a multiple of three, the ground state is unique, which is a peculiar feature for topologically ordered phases. 
	Second, while the ground state degeneracy depends on $L_x$ and $L_y$, it does not scale with them. It is finite in the thermodynamic limit. This is in accordance with the bound of the ground state degeneracy for homogeneous topological ordered phases obtained in \cite{Haah_2021}, $	\log(\text{GSD}) \sim L^{d-2}$, in $d$ spatial dimensions.


	\section{Symmetry Operators}\label{symmetryperators}
	
	Extended symmetry operators are intimately connected with the global constraints discussed in the previous section. They can be constructed by considering constraints in limited regions of the lattice, so that the violations of the constraints at the borders of the region give rise  to symmetry operators.

	\subsection{Horizontal Symmetry Operators}

	Let us start the discussion with $L_y$ being even. In this case, the relation \eqref{123} splits into constraints in even and odd sublattices. Then we consider \eqref{generalconstraint} but with the product in $y$-direction limited to a finite interval. Specifically, we take the product between the sites $n_1$ and $n_2$ of the even sublattice, 
	\begin{equation}
		\prod_{i_x=1}^{L_x} \prod_{\underset{i_y\, \in \, \text{even}}{i_y=n_1}}^{n_2} \left(P_{(i_x,i_y)}\right)^{t_{(i_x,i_y)}}. 
		\label{001}
	\end{equation}
	Naturally, this will no longer reduce to the identity since the product over $i_y$ is taken in an open region. The contributions from the interior of the region reduce to the identity in \eqref{001}, but the terms of the borders survive,
	\begin{equation}
		\prod_{i_x=1}^{L_x} \prod_{\underset{i_y\, \in \, \text{even}}{i_y=n_1}}^{n_2}\left(P_{(i_x,i_y)}\right)^{t_{(i_x,i_y)}} = \left( \prod_{i_x=1}^{L_x} (X_{(i_x, n_1-1)} )^{t_{(i_x,n_1)}} \right) \left( \prod_{i_x=1}^{L_x} (  X_{(i_x, n_2+1)})^{t_{(i_x,n_2)}} \right).
		\label{0012}
	\end{equation}
	As the left hand side of this expression is a product of commuting projectors, it commutes with the Hamiltonian. Now, in the right hand side, as the distance between the sites $n_1$ and $n_2$ can be arbitrarily large, and as the Hamiltonian is local, each one of the terms in the brackets must commute individually with the Hamiltonian, i.e., 
	\begin{equation}
		\prod_{i_x=1}^{L_x} (X_{(i_x, n_1-1)} )^{t_{(i_x,n_1)}} ~~~\text{and}~~~  \prod_{i_x=1}^{L_x} (  X_{(i_x, n_2+1)})^{t_{(i_x,n_2)}},
	\end{equation}
	are symmetry operators. However, they are not independent, as they can be deformed into each other through the product of commuting projectors, as specified in \eqref{0012}. This leads to a constraint between these two extended operators when projected in the ground state, where all the commuting projectors have eigenvalue equal to one.
	
	We then have the symmetry operators in the even and odd sublattices,  
	\begin{equation}
		W_{i_y\,\in\,\text{even}}\equiv\prod_{i_x=1}^{L_x} (X_{(i_x, i_y)} )^{t_{(i_x,i_y+1)}}
		\label{op1a}
	\end{equation}
	and
	\begin{equation}
		W_{i_y\,\in\,\text{odd}}\equiv\prod_{i_x=1}^{L_x} (X_{(i_x, i_y)} )^{t_{(i_x,i_y+1)}}.
		\label{op2b}
	\end{equation}
	When $L_y$ is odd, it turns out that the two above operators are not independent since starting in the even sublattice, we can reach the odd one through the application of commuting projectors around the $y$ direction. In addition, if $L_x$ is not a multiple of three, the only solution for the $t_{(i_x,i_y)}$ is the trivial one, and the operators \eqref{op1a} and \eqref{op2b} become trivial.
	
	\begin{figure}
		\centering
		\includegraphics[angle=90,scale=0.65]{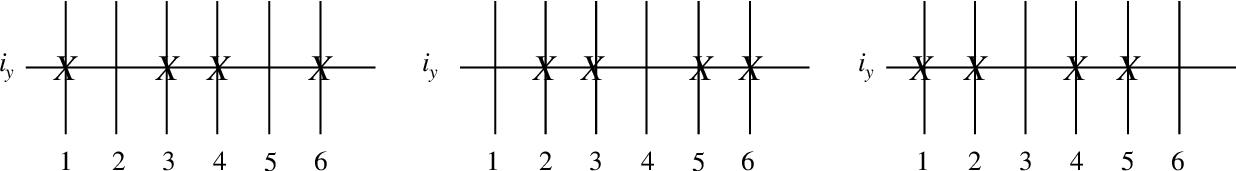}
		\caption{Horizontal symmetry operators with $i_y\,\in\, \text{even}$.}
		\label{op21}
	\end{figure}

	For $L_x$ being a multiple of three, the operators  \eqref{op1a} and \eqref{op2b} are nontrivial, and are of the form shown in Fig. \ref{op21}. For future purposes, we will specify such operators according to the notation
	\begin{equation}
		W_{i_y\,\in\,\text{even}}(n_{i_x-1},n_{i_x})~~~\text{and}~~~W_{i_y\,\in\,\text{odd}}(n_{i_x-1},n_{i_x}),
	\end{equation}
	making reference to the occupation numbers $(n_{i_x-1},n_{i_x})$\footnote{Note that this specification is sufficient to determine the whole extended operator.}  associated with an arbitrary pair of neighboring sites  $i_x-1$ and $i_x$, respectively. Specifically,  $n_{i_x}$ determines the existence or not of an $X$-operator acting at the site $i_x$ depending whether $n_{i_x}=0$ or 1. For example, the operators of Fig. \ref{op21} can be unambiguously specified by  
	\begin{equation}
		W_{i_y\,\in\,\text{even}}(n_{4}=1,n_{5}=0),~~~ W_{i_y\,\in\,\text{even}}(n_{4}=0,n_{5}=1),~~~W_{i_y\,\in\,\text{even}}(n_{4}=1,n_{5}=1),
	\end{equation}
	respectively. We also see that 
	\begin{equation}
		W_{i_y\,\in\,\text{even}}(n_{4}=1,n_{5}=0) W_{i_y\,\in\,\text{even}}(n_{4}=0,n_{5}=1)=W_{i_y\,\in\,\text{even}}(n_{4}=1,n_{5}=1),
	\end{equation}
	i.e., they are not independent. We then have two independent horizontal symmetry operators for the even sublattice and, similarly, two operators for the odd sublattice, totaling four symmetry operators. If $L_y$ is odd, the operators in the two sublattices are no longer independent, and we have only two operators in total.

\begin{figure}
		\centering
		\includegraphics[angle=90,scale=0.65]{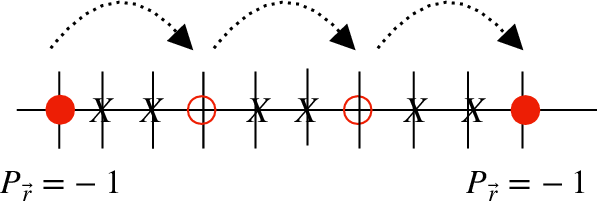}
		\caption{The filled red balls indicate the initial and final positions of an excitation in a transport process. The picture shows that an excitation can move along the $x$-direction only in rigid steps of size three in units of the lattice spacing. }
		\label{transportx}
	\end{figure}

	{\it Open} horizontal operators can be used to transport excitations\footnote{Transport processes involve open operators, whose existence does not depend whether $L_x$ is a multiple of three or not. }. The structure of the operator enforces the excitations to move along $x$-direction only in rigid steps of size three in units of the lattice spacing, as is illustrated in Fig. \ref{transportx}. This peculiarity of transport leads to the notion of position-dependent anyons. In a region of size three units of the lattice spacing, we can have three excitations but only two of them are independent (in the sense that given two excitations, the third one can be obtained through the application of a local operator - they belong to the same superselection sector. As the two excitations cannot be connected by transport, they actually correspond to distinct excitations, characterized by the position. When $L_x$ is not a multiple of three, it turns out that all the three excitations are in the same superselection sector, i.e., there is a single superselection sector, as there are extended operators creating  excitations in each of the positions.

	
	\subsection{Vertical Symmetry Operators}
	
\begin{figure}
		\centering
		\includegraphics[angle=90,scale=0.6]{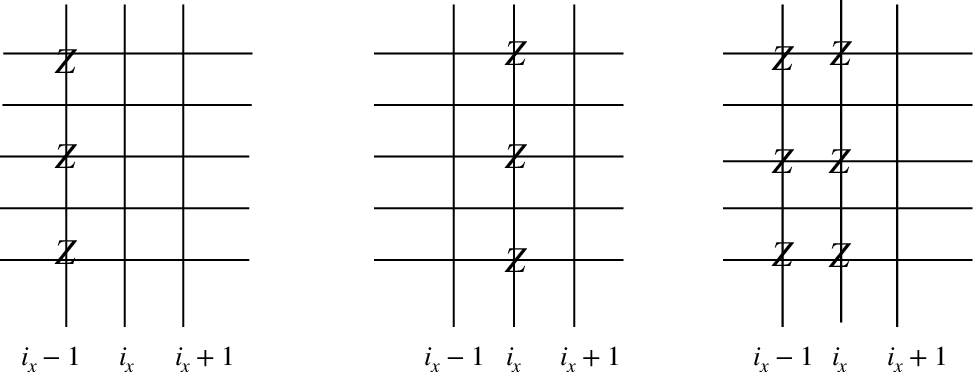}
		\caption{Vertical symmetry operators in the even $i_y$ sublattice. }
		\label{op1}
	\end{figure}

	The vertical symmetry operators can be constructed by following the same strategy, but considering now constraints limited in the $x$-direction. We consider first the case of $L_y$ even and take \eqref{generalconstraint} limited in the region $m_1 < i_x< m_2$,
	\begin{equation}
		\prod_{i_x=m_1}^{m_2}   \prod_{i_y\,\in\, \text{even}} \left(P_{(i_x,i_y)}\right)^{t_{(i_x,i_y)}}.
	\end{equation}
	All the contributions from the interior of the region reduce to the identity and we obtain
	\begin{eqnarray}
		\prod_{i_x=m_1}^{m_2}   \prod_{i_y\,\in\, \text{even}} \left(P_{(i_x,i_y)}\right)^{t_{(i_x,i_y)}}&=&  \left( \prod_{i_y\,\in\, \text{even}} (Z_{(m_1-1,i_y)})^{t_{(m_1,i_y)}}  (Z_{(m_1,i_y)})^{t_{(m_1,i_y)}+t_{(m_1+1,i_y)}}  \right)\nonumber\\
		&\times&  \left( \prod_{i_y\,\in\, \text{even}}  (Z_{(m_2,i_y)})^{t_{(m_2-1,i_y)}+t_{(m_2,i_y)}}  (Z_{(m_2+1,i_y)})^{t_{(m_2,i_y)}}   \right).
	\end{eqnarray}
	By the same argument as in the case of horizontal operators,  each one of the terms in the brackets must commute individually with the Hamiltonian, corresponding to symmetry operators  
	\begin{equation}
		\Gamma_{i_x}^{\text{even}} \equiv\prod_{i_y\,\in\, \text{even}} (Z_{(i_x-1,i_y)})^{t_{(i_x,i_y)}}  (Z_{(i_x,i_y)})^{t_{(i_x,i_y)}+t_{(i_x+1,i_y)}}. 
	\end{equation}
	The explicit form of these operators are shown in Fig. \ref{op1}. We also have symmetry operators in the odd sublattice, 
	\begin{equation}
		\Gamma_{i_x}^{\text{odd}} \equiv\prod_{i_y\,\in\, \text{odd}} (Z_{(i_x-1,i_y)})^{t_{(i_x,i_y)}}  (Z_{(i_x,i_y)})^{t_{(i_x,i_y)}+t_{(i_x+1,i_y)}}. 
	\end{equation}

	We will specify such operators by
	\begin{equation}
		\Gamma^{\text{even}}_{i_x}(m_{i_x-1},m_{i_x})~~~\text{and}~~~\Gamma^{\text{odd}}_{i_x}(m_{i_x-1},m_{i_x}),
		\label{verticaloperators}
	\end{equation}
	where $m_{i_x}=0,1$ indicates whether a line of $Z$-operators crosses the site  $i_x$. With this notation, the three operators in Fig. \ref{op1} are represented by 
	\begin{equation}
		\Gamma^{\text{even}}_{i_x}(1,0),~~~ \Gamma^{\text{even}}_{i_x}(0,1),~~~\Gamma^{\text{even}}_{i_x}(1,1),
	\end{equation} 
	and satisfy
	\begin{equation}
		\Gamma^{\text{even}}_{i_x}(1,0)\Gamma^{\text{even}}_{i_x}(0,1) = \Gamma^{\text{even}}_{i_x}(1,1),
	\end{equation}
	which shows that they are not independent.
	
	If $L_y$ is odd, all the $t_{\vec{r}}$'s along a line in the $y$-direction are constrained and therefore the vertical symmetry operators correspond to the product of the operators in \eqref{verticaloperators},  $\Gamma^{\text{even}}_{i_x}(m_{i_x-1},m_{i_x})\Gamma^{\text{odd}}_{i_x}(m_{i_x-1},m_{i_x})$.

	{\it Open} vertical operators are responsible for the mobility of excitations along the $y$-direction. The structure of the transport operators imply that the excitations can move only in rigid steps of size two units of the lattice spacing, i.e., in even or odd sublattices.  This also leads to the notion of position-dependent anyons along the $y$-direction.  If $L_y$ is odd, the excitations in even and odd sublattices belong to the same superselection sector.

	\subsection{'t Hooft Anomalies and Ground State Degeneracy}\label{thooftsection}
	
	The ground state degeneracy \eqref{GSD} can be understood in terms of 't Hooft anomalies between horizontal and vertical symmetry operators. 't Hooft anomalies are similar to projective representations in quantum mechanics, where certain operators acting in the Hilbert space fail to commute due to the existence of phases. This implies that the ground state cannot correspond to a one-dimensional representation, i.e., it cannot be unique. This can be rephrased in terms of a gauge obstruction, namely, the presence of a 't Hooft anomaly causes an obstruction to simultaneously gauging the corresponding symmetries \cite{tHooft:1979rat,Kapustin:2014lwa,Kapustin:2014zva}.

	Now we can analyze the commutation relations between the symmetry operators to check the existence of t'Hooft anomalies. As discussed previously, not all horizontal operators are independent, nor are all vertical operators. It is convenient to choose the following set of independent operators: $W_{i_y\,\in\,\text{even}}(1,0)$,  $W_{i_y\,\in\,\text{even}}(0,1)$, $\Gamma^{\text{even}}_{i_x}(1,0)$, and $\Gamma^{\text{even}}_{i_x}(0,1)$. We recall that the horizontal operators are fully specified by an arbitrary pair of neighboring sites 	$W_{i_y\,\in\,\text{even}}(n_{i_x-1},n_{i_x})$. Thus, in the commutation relation with vertical operator, we choose these sites to be precisely the ones entering $\Gamma^{\text{even}}_{i_x}(m_{i_x-1},m_{i_x})$, since these are the only sites relevant for the commutation relations.
	
	The set of independent operators satisfies
	\begin{eqnarray}
		&&\{W_{i_y\,\in\,\text{even}}(1,0), \Gamma^{\text{even}}_{i_x}(1,0)\}=0, ~~~
		\{W_{i_y\,\in\,\text{even}}(0,1), \Gamma^{\text{even}}_{i_x}(0,1)\} = 0,\nonumber\\
		&&~[W_{i_y\,\in\,\text{even}}(1,0), \Gamma^{\text{even}}_{i_x}(0,1)]=0, ~~~ [W_{i_y\,\in\,\text{even}}(0,1), \Gamma^{\text{even}}_{i_x}(1,0)]=0.
		\label{algebra}
	\end{eqnarray}
	The nontrivial commutation relations in the first line of this expression signalize a mixed 't Hooft anomaly between horizontal and vertical operators, leading to a ground state multiplicity of $2\times 2 =2^2$. For the case where $L_y$ is even, we also have the contribution from the operators in the odd sublattice, giving the total ground state degeneracy
	\begin{equation}
		\text{GSD} = 2^{4},
	\end{equation}
	which matches \eqref{GSD}. 
	For the cases where $L_x$ is not a multiple of three, the ground state is unique, since there are no nontrivial symmetry operators.

	
	\section{Local Indistinguishability}\label{li00}
	
	Topological order can be defined in terms of local indistinguishability of ground states \cite{Bonderson_2013},
	\begin{equation}
		\bra{GS,a} O(x) \ket{GS,b}= C \delta_{ab},
		\label{LI}
	\end{equation} 
	where the indices $a,b$ run over the degenerate ground states, $O(x)$ is a generic {\it local} operator, and $C$ is a constant. Relation \eqref{LI} encodes two important informations: i) the ground states cannot be distinguished by any local operator in the sense that $\bra{GS,1} O(x) \ket{GS,1}=\bra{GS,2} O(x) \ket{GS,2}=\cdots$\,; and ii) there is no local operator connecting different ground states, i.e., $\bra{GS,a} O(x) \ket{GS,b}=0$ for $a\neq b$.
	
	It turns out that  the mixed 't Hooft anomalies in the first line of \eqref{algebra} are sufficient to ensure the condition of local indistinguishability of ground states \eqref{LI}. To see this, let us pick up a pair of operators satisfying a nontrivial commutation relation like in \eqref{algebra},
	\begin{equation}
		\{W,\Gamma\}=0.
		\label{simplified}
	\end{equation}
	For simplicity we are omitting all the specifications that are not relevant for the discussion. This relation leads to a two-fold multiplicity of ground states. Suppose we choose to diagonalize the operator $W$, 
	\begin{equation}
		W \ket{GS,a} = \lambda_a \ket{GS,a}, ~~~a=1,2~~~ \text{and}~~~ \lambda_{1,2}=\pm 1.   
	\end{equation}
	Accordingly, we have
	\begin{equation}
		\Gamma \ket{GS,1} =  \ket{GS,2}.
		\label{003}
	\end{equation}
	This relation allows us to rephrase topological order in terms of spontaneous symmetry breaking, in the sense that distinct ground states are connected  by an {\it extended} symmetry operator (rather than a local one). In other words, certain extended symmetry oprators  act nontrivially on the ground state. This is precisely the notion of spontaneous symmetry breaking, but of a generalized symmetry.
	
	Now we proceed to show \eqref{LI}. Consider first the matrix element $\bra{GS,1} O(x) \ket{GS,2}$. By introducing the identity  $W^2=\openone$, it can be rewritten as 
	\begin{eqnarray}
		\bra{GS,1} O(x) \ket{GS,2} &=& \bra{GS,1} W^2 O(x)  \ket{GS,2}\nonumber\\
		&=&  \bra{GS,1} W O(x) W \ket{GS,2}\nonumber\\
		&=& \lambda_{1} \lambda_{2} \bra{GS,1} O(x) \ket{GS,2}.
		\label{0031}
	\end{eqnarray} 
	Note that from the first to the second line we have commuted the horizontal operator $W$ with $O(x)$, since the extended operators can always be deformed (through the multiplication by commuting projectors $P_{\vec{r}}$'s) in order to avoid the position $x$ of the local operator $O(x)$. Relation \eqref{0031} implies 
	\begin{equation}
		\bra{GS,1} O(x) \ket{GS,2}=0.
	\end{equation}
	
	Next we consider the expectation value $\bra{GS,2} O(x) \ket{GS,2}$. According to \eqref{003}, it can be written as
	\begin{eqnarray}
		\bra{GS,2} O(x) \ket{GS,2}&=& \bra{GS,1} \Gamma O(x) \Gamma \ket{GS,1}\nonumber\\
		&=& \bra{GS,1} O(x) \ket{GS,1},
		\label{0032}
	\end{eqnarray}
	which shows \eqref{LI}. In \eqref{0032}, we have commuted the extended operator $\Gamma$ with the local operator $O(x)$, since it can also be deformed to avoid the position of the local operator.
	
	\begin{figure}
		\centering
		\includegraphics[angle=90,scale=0.6]{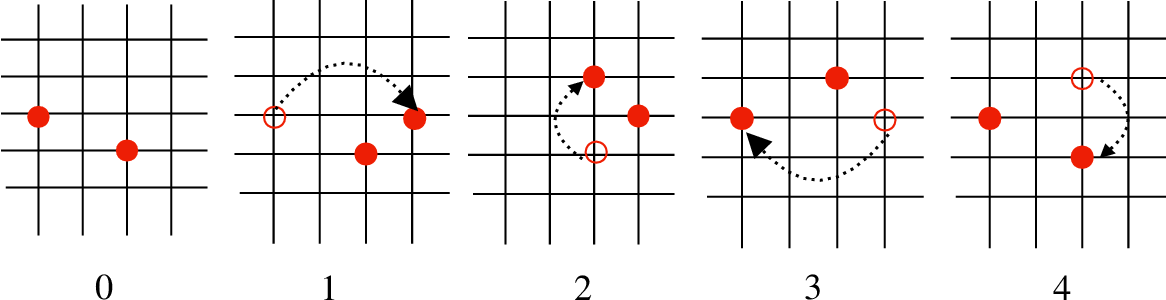}
		\caption{Nontrivial mutual statistics}
		\label{mutual}
	\end{figure}
	
	When $L_x$ is not a multiple of three, the ground state is unique and the condition \eqref{LI} does not bring much information, since it is trivially satisfied. In this case, we can understand that the system is topological ordered in terms of nontrivial mutual statistics between quasi-particles. We simply consider a two-particle state $\ket{\psi_2}$, as specified by the configuration 0 in Fig. \ref{mutual}. After the sequence of steps from 1 to 4, the two-particle state returns to its initial configuration leaving behind a nontrivial phase. In algebraic terms, the sequence of steps in Fig. \ref{mutual} is described by
	\begin{equation}
		\Gamma\,W\,\Gamma \,W\ket{\psi_2} = - \ket{\psi_2},
	\end{equation}
	where $W$ and $\Gamma$ are horizontal and vertical open operators satisfying $\{\Gamma, W\}=0$. The mutual statistics implies long-range entanglement, which is a signature of topological ordering. Note that this reasoning is independent of the linear sizes $L_x$ and $L_y$.

	
	\subsection{Topological Entanglement Entropy}\label{entanglement}
	
	An alternative way to understand that the system is topologically ordered is through the entanglement entropy. For a subregion $R$ of the system, the von Neumann entropy $S_R$ of a gapped system with local interactions exhibits the ``area law" behavior, 
	\begin{equation}
		S_R = \alpha L_R - S_{\text{topo}},
	\end{equation}
	where $\alpha$ is a nonuniversal constant, $L_R$ is the perimeter of the subregion $R$, and $S_{\text{topo}}$ is a universal constant reflecting topological ordering  \cite{Kitaev:2005dm,Levin_2006}. 
	
\begin{figure}
		\centering
		\includegraphics[scale=.3]{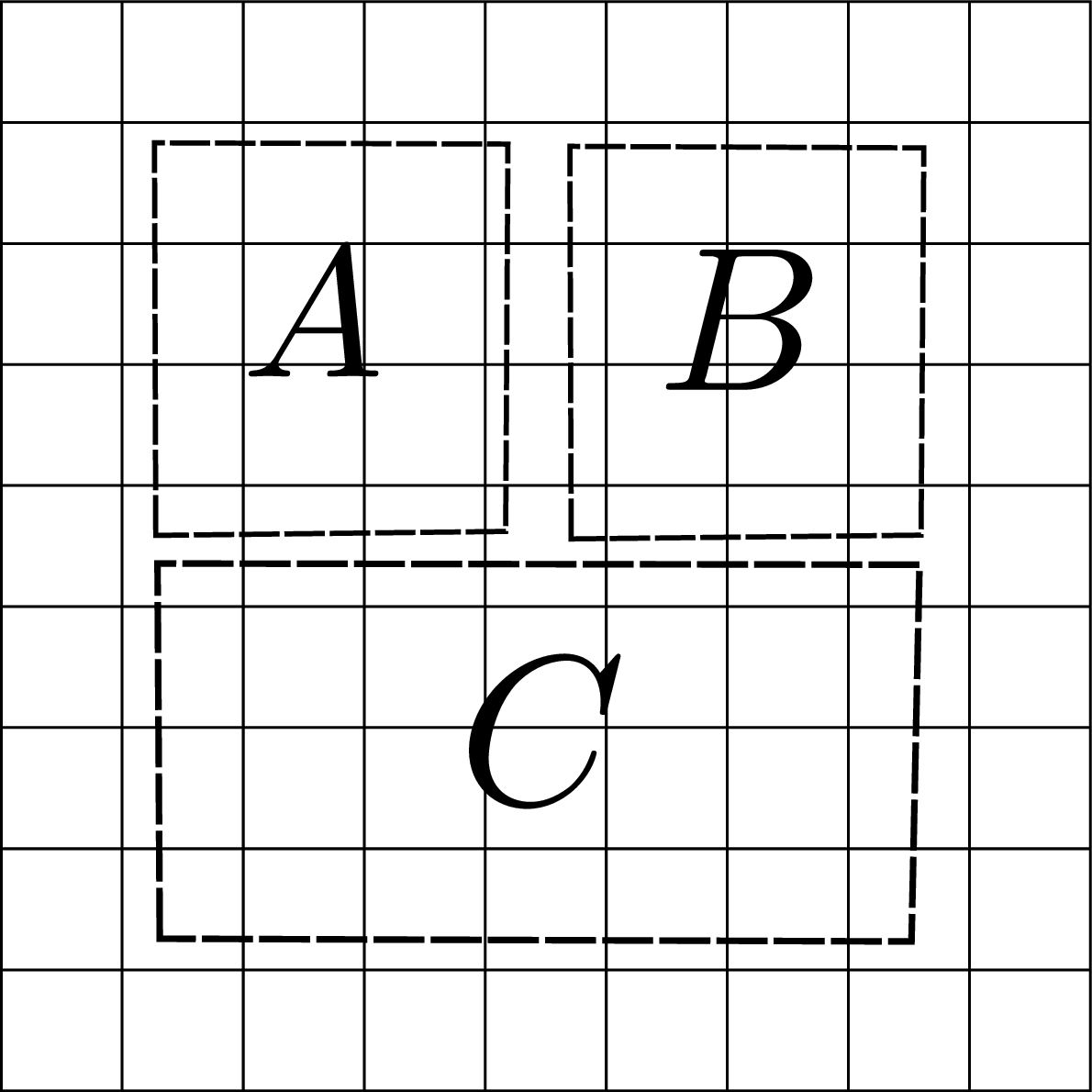}
		\caption{Division of the lattice in regions to calculate topological entropy.}
		\label{fig:region-division}
	\end{figure}

	The Kitaev-Preskill prescription \cite{Kitaev:2005dm} enables to isolate the topological part $S_{\text{topo}}$ in a simple way.  We split the subregion $R$ into three parts $A$, $B$, and $C$, as shown in Fig. \ref{fig:region-division}. Then, we compute the entropy in different partitions of $R$, and combine them as  
	\begin{equation}
		S_{\text{topo}} = S_{AB} + S_{BC} + S_{AC}  - S_A-S_B-S_C - S_{ABC}.
		\label{topo}
	\end{equation}
	This combination eliminates all the contributions from short-range entanglement (perimeter dependence) in the entropy, remaining only the topological contribution.

	Now we proceed to compute the entropy for an arbitrary subregion of the lattice, which is simple for a lattice model given in terms of commuting projectors. We follow the discussions in  \cite{linden_et_al:LIPIcs.TQC.2013.270,Zou:2016dck,Watanabe:2022pgk}. 
	
	Given a state $\ket{\psi}$, the von Neumann entropy for a subregion $R$ of the system is given by
	\begin{equation}
		S_R = -\Tr(\rho_R \log \rho_R),
		\label{3}
	\end{equation}
	where $\rho_R $ is the reduced density matrix $\rho_R \equiv\Tr_{\bar{R}} \ket{\psi}\bra{\psi}$, with the trace $\Tr_{\bar{R}}$ being taken over the region $\bar{R}$, which is the complement of $R$. Now, we consider a group $G$ given by all stabilizers $P_{\vec{r}}$ and their products, and also including simultaneously commuting extended symmetry operators.  Then we consider a particular ground state $\ket{\Phi_0}$, as the unique eigenstate of all elements $g \in G$ with eigenvalue $+1$. Such state can be expressed as $\ket{\Phi_0} = 1/\vert G \vert  \sum_{g} g \ket{\phi}$, where $\ket{\phi}$ is a reference state. Consequently, the density matrix can be written as
	\begin{equation}
		\rho = \frac{1}{\vert G \vert} \sum_{g\, \in\, G} g,
	\end{equation}
	where $\vert G \vert$ is the number of elements in $G$.

	To compute the entropy \eqref{3} for the state $\ket{\Phi_0}$, we need to take the trace over the region $\bar{R}$,
	\begin{equation}
		\rho_R = \frac{1}{\vert G \vert} \sum_{g\, \in\, G} \Tr_{\bar{R}}(g).
	\end{equation}
	Since $g$ is the product of Pauli matrices and identities, this trace is nonvanishing only when $g$ acts as the identity over the region $\bar{R}$, so that the sum reduces to the subgroup $G_R$, which is the group of elements of $G$ that are identities over $\bar{R}$ (they act nontrivially only in the region $R$).
	The trace gives the dimension of the Hilbert space restricted to the region $\bar{R}$, which is $2^{n_{\bar{R}}}$. This gives the reduced density matrix over $R$,
	\begin{equation}
		\rho_R = \frac{1}{2^{n_R}} \sum_{g \,\in \,G_R} g,
	\end{equation}
	where $n_R$ is the number of spins in the region $R$. The reduced density matrix is proportional to the projector
	\begin{equation}
		P_{G_R}\equiv\frac{1}{|G_R| } \sum_{g \,\in \,G_R} g,
	\end{equation}
	where $|G_R|$ is the number of elements in $G_R$.  Thus, the only nonvanishing eigenvalue of $\rho_R$ is  ${|G_R|}/{2^{n_R}}$, with degeneracy given by the number of spins in the region $R$ divided by the order of the subgroup $G_R$, namely, $2^{n_R}/|G_R|$. 
	
	With this, after computing the trace in (\ref{3}), we find the simple expression
	\begin{equation}
		S_R = n_R \log 2 - \log \vert G_R \vert.
		\label{ee01}
	\end{equation}
	To proceed, we will take specifically the regions $A$, $B$ and $C$ of Fig. \ref{fig:region-division}, but the calculation is independent of the size of the regions. We need to determine how many commuting projectors there are in each one of the regions. In $A$ and $B$ there is only one stabilizer; in $C$ there are 4. In $AB$, there are 4; in $AC$ and $BC$ there are $8$. Finally, in $ABC$ there are 16. Collecting all contributions according to \eqref{topo}, we find a nontrivial topological contribution
	\begin{equation}
		S_{\text{topo}} =  52 \log 2 -30 \log 2  - 20 \log 2 =  \log 4,
		\label{entropy0}
	\end{equation}
	indicating that the ground state is  topologically ordered. 
	
	The topological contribution can also be expressed in terms of the total quantum dimension of anyons
	\begin{equation}
		\mathcal{D}=\sqrt{\sum_a d_a^2},
	\end{equation}
	with $d_a$ being the quantum dimension of the anyon $a$. In terms of $\mathcal{D}$, the topological entropy reads \cite{Kitaev:2005dm}
	\begin{equation}
		S_{\text{topo}}=\log \mathcal{D}.
		\label{990099}
	\end{equation}
	For the present case of Abelian anyons, where $d_a=1$, the result \eqref{entropy0} matches with the particle spectrum of 16 anyons,  following from all possible combinations of the maximum number of basic excitations, which is four (two in the horizontal even and two in the horizontal odd sites). The above computation, however, disregards the region outside $A$, $B$, and $C$, and consequently does not take into account that for certain sizes it may happen that all  anyons belong to the same superselection sector, for example, when $L_x$ is not a multiple of three, so that we have in effect a single excitation.
	
	A refinement of the entanglement entropy computation, which reflects the sensitivity of the number of superselection sectors (which in turn depends on the lattice sizes) is discussed in Ref. \cite{Kim:2023qqi}. The idea is to consider the system in a torus and then to compute the topological contribution for a non-contractible region of the torus, i.e., a region which winds around the torus along one direction. This analysis is carried out in the appendix \ref{appendixB}.



	\section{Boundary Theory}\label{boundarytheory0}
	
	A topological phase with symmetry pattern $\mathbb{Z}_2 \times \mathbb{Z}_2$ exhibits a rich phase structure, which will be explored in this section.

	\subsection{Ground State Degeneracy and Boundary Symmetry Operators}
	
	In the presence of boundaries, there are no longer global constraints and also there is no longer a one-to-one correspondence between sites and commuting projectors. The ground state degeneracy in this case scales exponentially with the boundary size. For example, for a system limited in the vertical direction by two boundaries with a total of $L_x L_y$ sites, the number of operators in the Hamiltonian is $L_x L_y - 2L_y$, since each one of the vertical boundaries involves $L_y$ sites which do not have an associated commuting projector (because the commuting projector does not fit into the system if it is on the boundary). Thus, the ground state degeneracy is 
	\begin{equation}
		\text{GSD}_{\text{boundary}}= \frac{2^{L_xL_y}}{2^{L_x L_y-2L_y}} = 2^{2L_y},
		\label{ghjk}
	\end{equation}
	which scales exponentially with the boundary length. Similarly, a system limited in the horizontal direction by two boundaries has the ground state degeneracy
	\begin{equation}
		\text{GSD}_{\text{boundary}}= \frac{2^{L_xL_y}}{2^{L_x L_y-2L_x}} = 2^{2L_x}.
		\label{ghjk1}
	\end{equation}
	Such large degeneracies \eqref{ghjk} and  \eqref{ghjk1} mean that there must be an extensive number of anomalies at the boundaries. The origin of such anomalies is the existence of boundary symmetry operators.

\begin{figure}
		\centering
		\includegraphics[angle=90,scale=0.65]{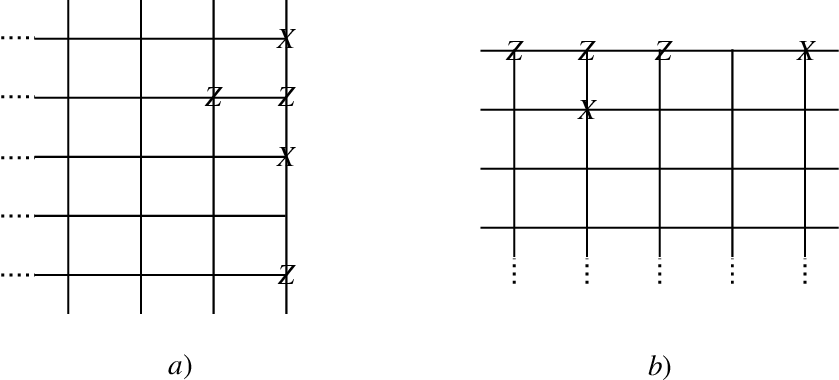}
		\caption{a) Boundary operators for a vertical boundary at the right. b) Boundary operators for a horizontal boundary at the top. }
		\label{boundaryops}
	\end{figure}

	When we consider open boundary conditions we have new operators defined at the boundaries of the lattice that commute with the stabilizers of the bulk Hamiltonian. These new operators arise by considering incomplete version of the  bulk stabilizers. For vertical and horizontal cuts shown in Fig. \ref{boundaryops}, they are
	\begin{align}
		B_{1, i_y}^{v}=\, \begin{matrix}
			{}&X
			\\
			Z&Z_{i_y}
			\\
			{} & X
		\end{matrix}
		\qquad\qquad\text{and} \qquad\qquad B_{2, i_y}^{v}=Z_{i_y}, 
		\label{stab vertical}
	\end{align}
	and
	\begin{align}
		B^{h}_{1, i_x}=\,\begin{matrix}
			Z & Z_{i_x} & Z \\
			{} & X &{}	
		\end{matrix}
		\qquad\qquad\text{and} \qquad\qquad B_{2,i_x}^{h}=X_{i_x},
		\label{stab horizontal}
	\end{align}
	respectively. For cuts opposite to those in the Fig. \ref{boundaryops}, the corresponding boundary operators are mirror reflected.

	\subsection{Anomaly Counting for Vertical Boundaries}
	
	With the boundary symmetry operators we can explicitly count the number of anomalies. Let us study now the case of vertical boundaries.
	
	Note that the boundary operators  \eqref{stab vertical} satisfy
	\begin{equation}
		\{B^v_{1,i_y}, B^v_{2,i_y\pm1} \}=0,
	\end{equation}
	i.e., each operator $B^v_{1,i_y}$ anticommutes with two other operators. 
	In addition, the vertical boundary operators generically do not commute with $W_{i_y}(n_{i_x-1},n_{i_x})$. This mixing of noncommuting operators makes it difficult to count the number of independent anomalies. The idea is to find a new basis (canonical) for the boundary operators where the anticommutation relations are pairwise decoupled. This can be done in the following way.
	
	Starting with the case of $L_y$ even, the set of symmetry operators is
	\begin{equation}
		\{W_{i_y\,\in\,\text{even}/\text{odd}}(n_{i_x-1},n_{i_x}),\Gamma^{\text{even}/\text{odd}}_{i_x}(m_{i_x-1},m_{i_x}), B^v_{1,i_y}, B^v_{2,i_y}\}.
		\label{setofop}
	\end{equation}
	Notice, however, that not all operators in this set are independent. We already know that only four $W'$s and four $\Gamma$'s are independent (for the same reason as in the case of periodic boundary conditions). In addition, the boundary operators are not all independent, as we can express products of boundary operators in terms of $\Gamma^{\text{even}/\text{odd}}$:
	\begin{equation}
		\prod_{i_y \,\in\,\text{even}} B^v_{1,i_y} =  \Gamma^{\text{even}}_{L_x}(1,1)~~~\text{and}~~~\prod_{i_y \,\in\,\text{even}} B^v_{2,i_y} =  \Gamma^{\text{even}}_{L_x}(0,1),
	\end{equation}
	and 
	\begin{equation}
		\prod_{i_y \,\in\,\text{odd}} B^v_{1,i_y} =  \Gamma^{\text{odd}}_{L_x}(1,1)~~~\text{and}~~~\prod_{i_y \,\in\,\text{odd}} B^v_{2,i_y} =  \Gamma^{\text{odd}}_{L_x}(0,1).
	\end{equation}	
	Thus, from all $B_{1}^v$'s, a number $L_y -2$ of them are in fact independent, and the same number holds for the operators $B_{2}^v$'s.	With this, from the set \eqref{setofop}, we exclude, say,  $B_{1,L_y-1}^v$, $B_{1,L_y}^v$, $B_{2,L_y-1}^v$, and $B_{2,L_y}^v$.  With the remaining $B_{1}^v$'s operators, we take the products 
	\begin{equation}
		\tilde{B}_{1,i_y\,\in\,\text{odd}}^v \equiv \prod_{\underset{j_y\, \in \, \text{even}}{j_y=i_y+1}}^{L_y-2} B_{1,j_y}^v,~~~ i_y=1,3,\ldots,L_y-3.
		\label{uio}
	\end{equation}	
	and
	\begin{equation}
		\tilde{B}_{1,i_y\,\in\,\text{even}}^v \equiv \prod_{\underset{j_y\, \in \, \text{odd}}{j_y=1}}^{i_y-1} B_{1,j_y}^v,~~~ i_y=2,4,\ldots,L_y-2.
		\label{uio1}
	\end{equation}		
	
\begin{figure}
		\includegraphics[angle=90,scale=0.65]{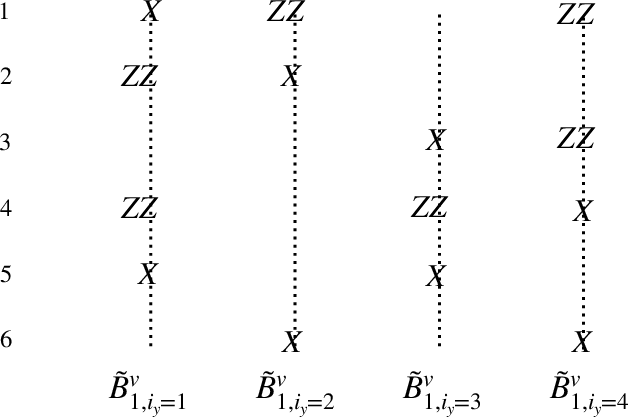}
		\caption{Examples of the operators constructed according to Eqs. \eqref{uio} and \eqref{uio1}.}
		\label{redops}
	\end{figure}

	An example of these operators for the size $L_y=6$ is shown in Fig. \ref{redops}. From the figure it is clear that the operators 
	\eqref{uio} and \eqref{uio1} do not involve $Z$'s in the sites $i_y=L_y-1$ and $i_y=L_y$. In this way, we can conveniently choose the four independent symmetry operators lying along horizontal lines crossing the sites $L_y-1$ and $L_y$, namely, we choose $W_{L_y-1}(n_{i_x-1},n_{i_x})$ and $W_{L_y}(n_{i_x-1},n_{i_x})$. This choice ensures that these operators commute with all independent boundary operators.

	The algebra of the symmetry operators is the following. The extended operators satisfy \eqref{algebra} and the boundary operators satisfy
	\begin{equation}
		\{\tilde{B}_{1,i_y}^v,B_{2,j_y}^v \}=\delta_{i_y,j_y},~~~ i_y, j_y=1,3,\ldots L_y-3
	\end{equation}	
	and 	
	\begin{equation}
		\{\tilde{B}_{1,i_y}^v,B_{2,j_y}^v \}=\delta_{i_y,j_y},~~~ i_y, j_y=2,4,\ldots L_y-2.
	\end{equation}		
	All other commutations are trivial. The above relations implies that the number of anomalies per boundary is $2\times(L_y/2-1)=L_y-2$.
	Thus, the total number of anomalies, taking into account the presence of two boundaries and the bulk, is 
	\begin{equation}
		\underbrace{4}_{\text{bulk anomalies}}+ \underbrace{(2L_y-4)}_{\text{boundary anomalies}} = 2L_y,
		\label{nm}
	\end{equation}
	which gives rise to the degeneracy \eqref{ghjk}.
	
	To conclude the discussion, we consider the case of $L_y$ odd. The discussion is similar to the previous case, but we have to take into account that the vertical symmetry operators involve the product in odd and even sublattices, 
	\begin{equation}
		\Gamma(m_{i_x-1},m_{i_x}) = \Gamma^{\text{odd}}_{i_x}(m_{i_x-1},m_{i_x})  \Gamma^{\text{even}}_{i_x}(m_{i_x-1},m_{i_x}),
	\end{equation}
	giving rise to only two vertical extended symmetry operators. Accordingly, the symmetries $W_{i_y}(n_{i_x-1},n_{i_x})$ need not to be split in even and odd sublattices. Consequently, there are two anomalies in the bulk.
	
	The boundary operators are not all independent due to the constraints
	\begin{equation}
		\prod_{i_y} B_{1,i_y}^v = \Gamma(1,1) ~~~ \text{and}~~~ \prod_{i_y} B_{2,i_y}^v = \Gamma(0,1),
	\end{equation}
	so that we have a number $L_y-1$ of independent boundary operators $B_{1}^v$'s and the same number of independent $B_{2}^v$'s.
	
	The final step is to construct a new basis for the boundary operators $B_1^v$'s so that the nontrivial commutations decouple pairwise. 
	The new basis can be constructed with a minor modification of \eqref{uio} and \eqref{uio1}, namely, as  
	\begin{equation}
		\tilde{B}_{1,i_y\,\in\,\text{odd}}^v \equiv \prod_{\underset{j_y\, \in \, \text{even}}{j_y=i_y+1}}^{L_y-1} B_{1,j_y}^v,~~~ i_y=1,3,\ldots,L_y-2.
		\label{uio2}
	\end{equation}	
	and
	\begin{equation}
		\tilde{B}_{1,i_y\,\in\,\text{even}}^v \equiv \prod_{\underset{j_y\, \in \, \text{odd}}{j_y=1}}^{i_y-1} B_{1,j_y}^v,~~~ i_y=2,4,\ldots,L_y-1.
		\label{uio3}
	\end{equation}		
	It is simple to check that the algebra of operators is
	\begin{equation}
		\{\tilde{B}_{1,i_y}^v,B_{2,j_y}^v \}=\delta_{i_y,j_y},~~~ i_y, j_y=1,2,\ldots L_y-1,
	\end{equation}	
	with these operatos commuting trivially with $\Gamma$ and $W$.

	The total number of anomalies (bulk + two boundaries) is 
	\begin{equation}
		\underbrace{2}_{\text{bulk anomalies}}+ \underbrace{(2L_y-2)}_{\text{boundary anomalies}} = 2L_y.
		\label{nm1}
	\end{equation}
	It is interesting to the see that \eqref{nm} and \eqref{nm1} can be unified as
	\begin{equation}
		\underbrace{N(L_y)}_{\text{bulk anomalies}}+~~ \underbrace{(2L_y-N(L_y))}_{\text{boundary anomalies}} = 2L_y,
	\end{equation}
	where $N(L_y)$ is the number of bulk anomalies depending on the size $L_y$, i.e., $N(L_y)=4$ for $L_y$ even and $N(L_y)=2$ for $L_y$ odd. In other words, there is a conservation of the total number of anomalies in the whole bulk-boundary system, in order to ensure the ground state degeneracy \eqref{ghjk}.
	
	The anomaly counting in the case of horizontal boundaries can be carried in a similar way.



	\begin{figure}
		\centering
		\includegraphics[angle=90,scale=0.65]{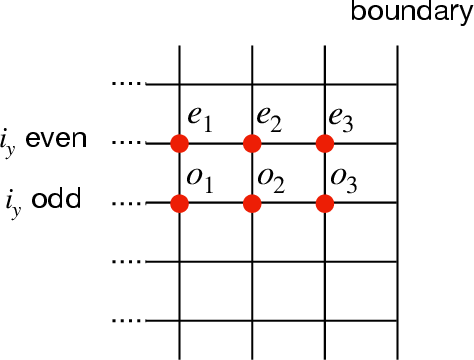}
		\caption{Trios of excitations in even and odd sublattices.  }
		\label{trio}
	\end{figure}
	
	\subsection{Boundary Phases}
	
	To produce stable gapped boundary phases we need to lift the large degeneracies in \eqref{ghjk} and \eqref{ghjk1}, which can be done by adding to the bulk Hamiltonian a set of simultaneously commuting boundary symmetry operators. As the boundary stabilizers may not commute with each other, each distinct set of commuting operators gives rise to a different phase, characterized by the condensation of certain excitations at the boundary. Naturally, condensation requires that the involved excitations have trivial mutual statistics. 
	
	We label a trio of neighboring excitations in the even and odd sublattices by $e_1, e_2, e_3$ and $o_1, o_2, o_3$, respectively. They are shown in Fig. \ref{trio}.
	As discussed previously, only two  excitations in each group are independent. Let us say that the independent excitations are $e_1$, $e_2$, $o_1$, and $o_2$.  The sets of individual excitations with trivial mutual statistics are
	\begin{equation}
		\{e_1, e_2\},~  \{o_1, o_2\},~ \{e_1, o_1\},~\{o_1, e_2\}. 
		\label{ind}
	\end{equation}
	In addition, we can have condensation of composite excitations that have mutual statistics, namely, 
	\begin{equation}
		\{e_1e_2, o_1o_2\},~ \{e_1o_1, e_2o_2\}.
		\label{comp}
	\end{equation}
	Therefore, we can construct boundary Hamiltonians corresponding to the condensation of each one of the six sets above. 
	
	Condensations are characterized by boundary Hamiltonians that commute with transport operators, otherwise the excitation can be localized at the boundary. For the excitations of Fig. \ref{trio}, the operators that transport such excitations to the boundary are
	\begin{eqnarray}
		&&e_1: ~~   X_{L_x-2,i_y \,\in\, \text{even}}\,X_{L_x-1,i_y \,\in\, \text{even}};\nonumber\\
		&& e_2: ~~ X_{L_x-1,i_y \,\in\, \text{even}}\,X_{L_x,i_y \,\in\, \text{even}}; \nonumber\\
		&& o_1:~~ X_{L_x-2,i_y \,\in\, \text{odd}} \,X_{L_x-1,i_y\, \in\, \text{odd}};\nonumber\\
		&& o_2:~~ X_{L_x-1,i_y \,\in\, \text{odd}}\, X_{L_x,i_y \,\in\, \text{odd}}.
	\end{eqnarray}
	The transport of composite excitations is done simply by the composition of the transport operators of the corresponding individual excitations.
	
	Given the anisotropic nature of both the Hamiltonian and the boundary stabilizers, we shall analyze the vertical and horizontal boundaries separately. In the following we study the case of vertical boundary and we leave for the appendix \ref{appendixA} the study of horizontal boundary.

	
	\subsection{Vertical Boundary}\label{Vertical Boundary}

	We construct here the boundary Hamiltonians corresponding to each one of the phases in \eqref{ind} and \eqref{comp}. In addition, we fully characterize the nature of the resulting phases according to their symmetries. In the whole discussion, we consider that $L_y$ is even, so that the vertical boundary operators are defined in even and odd sublattices.

	\subsubsection{Condensation of $\{e_1,o_1\}$ }
	
	Let us start with the Hamiltonian involving the boundary operator $B_2^v$,
	\begin{equation}
		H^v(e_1,o_1) =H_{\text{bulk}}\, - \sum_{i_y\,\in\, \text{even}} B_{2,i_y}^v - \sum_{i_y\,\in\,\text{odd}} B_{2,i_y}^v  =H_{\text{bulk}}\, - \sum_{i_y} Z_{i_y}.
		\label{eo1}
	\end{equation}
	The boundary contribution is just the Hamiltonian of an ideal paramagnet, which is trivially gapped. It is immediate to see from Fig. \ref{trio} that the operators which individually transport the excitations $e_1$ and $o_1$ to the boundary commute with the Hamiltonian \eqref{eo1}. Thus, these excitations are condensed at the boundary. On the other hand, the operators that transport the excitations $e_2$ and $o_2$ to the boundary do not commute the  \eqref{eo1} and, consequently, these excitations can be detected at the boundary.
	
	The bulk symmetry operators \eqref{verticaloperators} can be placed at the boundary,
	\begin{equation}
		\Gamma_{L_x}^{\text{even}}(0,1)~~~\text{and}~~~ \Gamma_{L_x}^{\text{odd}}(0,1),
		\label{jkl1}
	\end{equation}
	and remain as exact symmetry operators there.
	
	
	\subsubsection{Condensation of $\{e_2,o_2\}$ }
	
	Next we consider the boundary Hamiltonian constructed in terms of $B_1^v$ operators:
	\begin{equation}
		H^v(e_2,o_2)=H_{\text{bulk}} \, - \sum_{i_y\,\in\, \text{even}} B_{1,i_y}^v - \sum_{i_y\,\in\,\text{odd}} B_{1,i_y}^v =H_{\text{bulk}}\, -\sum_{i_y} \begin{matrix}
			{}&X
			\\
			Z&Z_{i_y}
			\\
			{} & X
		\end{matrix}.
		\label{eo2}
	\end{equation}
	The excitations $e_2$ and $o_2$ can be transported to the boundary and condensed there since the corresponding transport operators commute with the Hamiltonian \eqref{eo2}. 
	
	To characterize the boundary phase, we can analyze the properties of an effective boundary theory obtained by ``integrating out the bulk". Namely,  we consider the ground state of the bulk, so that in this situation the spins located at $(L_x-1,i_y)$, which appear in the boundary operators, are frozen. Thus, the effective boundary Hamiltonian reads
	\begin{equation}
		H_{\text{eff}}^v(e_2,o_2)=-\sum_{i_y} \begin{matrix}
			X
			\\
			Z_{i_y}
			\\
			X
		\end{matrix}.
		\label{eo2a}
	\end{equation}
	This is simply the cluster model \cite{2006RpMP...57..147N,Son_2011}. The ground state of \eqref{eo2a} is unique and it is protected by the symmetries
	\begin{equation}
		\Gamma_{L_x}^{\text{even}}(0,1)~~~\text{and}~~~ \Gamma_{L_x}^{\text{odd}}(0,1).
		\label{0543}
	\end{equation}
	If we place \eqref{eo2a} on a lattice with boundaries, the above symmetries are projectively realized at the boundary of the 1+1 dimensional system. Thus, \eqref{eo2a} describes a SPT phase.

	
	\subsubsection{Condensation of $\{o_1,o_2\}$ }
	
	The boundary phase that condenses the excitations $o_1$ and $o_2$ is described by 
	\begin{equation}
		H^v(o_1,o_2)=H_{\text{bulk}}\,  -\sum_{i_y\,\in\, \text{even}} (B_{1,i_y}^v + B_{2,i_y}^v )
		=H_{\text{bulk}}\, -\sum_{i_y\,\in\, \text{even}}  \begin{matrix}
			{}&X
			\\
			Z&Z_{i_y}
			\\
			{} & X
		\end{matrix}
		-\sum_{i_y\,\in\, \text{even}} Z_{i_y}. 
		\label{234}
	\end{equation}
	Proceeding as in the previous case, we obtain the effective boundary theory after integrating out the bulk, 
	\begin{equation}
		H_{\text{eff}}^v(o_1,o_2) = 	-\sum_{i_y\,\in\, \text{even}}  \begin{matrix}
			X
			\\
			Z_{i_y}
			\\
			X
		\end{matrix}
		-\sum_{i_y\,\in\, \text{even}} Z_{i_y}. 
		\label{234a}
	\end{equation}
	The symmetry operators are 
	\begin{equation}
		\Gamma_{L_x}^{\text{even}}(0,1)~~~\text{and}~~~ \Gamma_{L_x}^{\text{odd}}(0,1).
		\label{6758}
	\end{equation}
	In addition, there is a local operator that commutes with the Hamiltonian \eqref{234a}, 
	\begin{equation}
		X_{L_x, i_y\,\in\,\text{odd}}.
	\end{equation}
	However, this operator does not commute with $ \Gamma_{L_x}^{\text{odd}}(0,1)$, i.e.,
	\begin{equation}
		\{X_{L_x, i_y\,\in\,\text{odd}}, \Gamma_{L_x}^{\text{odd}}(0,1) \}=0.
	\end{equation}
	We can take a linear combination of the local operators,  
	\begin{equation}
		\Sigma_{\text{odd}}\equiv\frac{1}{L_y/2}\sum_{i_y\,\in\,\text{odd}} X_{L_x,i_y},
	\end{equation}
	which allows us to distinguish states. The symmetry operators satisfy
	\begin{equation}
		\{	\Sigma_{\text{odd}}, \Gamma_{L_x}^{\text{odd}}(0,1) \}=0~~~\text{and}~~~ [	\Sigma_{\text{odd}}, \Gamma_{L_x}^{\text{even}}(0,1)]=0.
	\end{equation}
	The nontrivial commutation relation between $\Sigma_{\text{odd}}$ and $\Gamma_{L_x}^{\text{odd}}(0,1)$ implies that the ground state is two-fold degenerate and hence this phase corresponds to a spontaneous symmetry-breaking phase.  The symmetry generated by $ \Gamma_{L_x}^{\text{even}}(0,1)$ is exact.
	
	To further appreciate this point, we can construct explicitly the ground states of \eqref{234a},
	\begin{equation}
		\bigotimes_{i_y\,\in\,\text{even}}
		\ket{ \begin{smallmatrix}
				+
				\\
				~~\uparrow_{i_y}
				\\
				+
		\end{smallmatrix}}
		~~~\text{and}~~~
		\bigotimes_{i_y\,\in\,\text{even}}
		\ket{ \begin{smallmatrix}
				-
				\\
				~~\uparrow_{i_y}
				\\
				-
		\end{smallmatrix}},
		\label{groundstates1}
	\end{equation}
	where $\ket{\pm}$ are eigenstates of $X$ and $\ket{\uparrow}$ is an eigenstate of $Z$. The notation in the kets is in accordance with the notation of the vertical operators of the Hamiltonian \eqref{234a}. Alternatively, we can take linear combinations of the ground states \eqref{groundstates1}.
	
	For the ground states as in \eqref{groundstates1}, we see that $ \Gamma_{L_x}^{\text{odd}}(0,1) $ acts nontrivially on them,
	\begin{equation}
		\Gamma_{L_x}^{\text{odd}}(0,1) 	\bigotimes_{i_y\,\in\,\text{even}}	\ket{ \begin{smallmatrix}
				+
				\\
				~~\uparrow_{i_y}
				\\
				+
		\end{smallmatrix}}= 	\bigotimes_{i_y\,\in\,\text{even}}	\ket{ \begin{smallmatrix}
				-
				\\
				~~\uparrow_{i_y}
				\\
				-
		\end{smallmatrix}},
		\label{groundstates}
	\end{equation} 
	while the local operator $	\Sigma_{\text{odd}}$ enables us to distinguish the degenerate ground  states,
	\begin{equation}
		\Sigma_{\text{odd}} 	\bigotimes_{i_y\,\in\,\text{even}}	\ket{ \begin{smallmatrix}
				+
				\\
				~~\uparrow_{i_y}
				\\
				+
		\end{smallmatrix}} = (+1)	\bigotimes_{i_y\,\in\,\text{even}}	\ket{ \begin{smallmatrix}
				+
				\\
				~~\uparrow_{i_y}
				\\
				+
		\end{smallmatrix}}
	\end{equation}
	and
	\begin{equation}
		\Sigma_{\text{odd}} 	\bigotimes_{i_y\,\in\,\text{even}}
		\ket{ \begin{smallmatrix}
				-
				\\
				~~\uparrow_{i_y}
				\\
				-
		\end{smallmatrix}}=(-1)	\bigotimes_{i_y\,\in\,\text{even}}
		\ket{ \begin{smallmatrix}
				-
				\\
				~~\uparrow_{i_y}
				\\
				-
		\end{smallmatrix}}.
	\end{equation}
	Therefore, the symmetry generated by $\Gamma_{L_x}^{\text{odd}}(0,1)$ is spontaneously broken, while the one generated by $\Sigma_{\text{odd}}$ is exact. 
	
	If we take linear combinations of the ground states in  \eqref{groundstates1}, 
	\begin{equation}
		\bigotimes_{i_y\,\in\,\text{even}}
		\left(\,
		\ket{ \begin{smallmatrix}
				+
				\\
				~~\uparrow_{i_y}
				\\
				+
		\end{smallmatrix}}
		\pm
		\ket{ \begin{smallmatrix}
				-
				\\
				~~\uparrow_{i_y}
				\\
				-
		\end{smallmatrix}}\,\right),
		\label{groundstates2}
	\end{equation}
	the realization of the symmetries generated by $\Sigma_{\text{odd}}$ and $\Gamma_{L_x}^{\text{odd}}(0,1)$ is interchanged, namely, the above ground states spontaneously break 
	the symmetry associated with $\Sigma_{\text{odd}}$, while the one associated with $\Gamma_{L_x}^{\text{odd}}(0,1)$ is exact.
	
	The operator $\Gamma_{L_x}^{\text{even}}(0,1)$  acts trivially on both ground states \eqref{groundstates1} and \eqref{groundstates2}, so that the corresponding symmetry is always exactly realized.

	
	\subsubsection{Condensation of $\{e_1,e_2\}$ }
	
	The boundary Hamiltonian that condenses $e_1$ and $e_2$ is
	\begin{equation}
		H^v(e_1,e_2)=H_{\text{bulk}}\, - \sum_{i_y\,\in\, \text{odd}} (B_{1,i_y}^v + B_{2,i_y}^v )
		=H_{\text{bulk}} \,-\sum_{i_y\,\in\, \text{odd}}  \begin{matrix}
			{}&X
			\\
			Z&Z_{i_y}
			\\
			{} & X
		\end{matrix}
		-\sum_{i_y\,\in\, \text{odd}} Z_{i_y}.
	\end{equation}
	The effective boundary Hamiltonian is
	\begin{equation}
		H_{\text{eff}}^v(e_1,e_2)=-\sum_{i_y\,\in\, \text{odd}}  \begin{matrix}
			X
			\\
			Z_{i_y}
			\\
			X
		\end{matrix}
		-\sum_{i_y\,\in\, \text{odd}} Z_{i_y}.
		\label{234a1}
	\end{equation}
	
	The analysis is essentially the same as in the previous case, with the role of even and odd symmetries interchanged. The symmetry operators are \eqref{6758}, but the local operator that commutes with the Hamiltonian is 
	\begin{equation}
		X_{L_x, i_y\,\in\,\text{even}}.
	\end{equation}
	This local operator does not commute with $\Gamma_{L_x}^{\text{even}}(0,1)$,
	\begin{equation}
		\{X_{L_x, i_y\,\in\,\text{odd}}, \Gamma_{L_x}^{\text{even}}(0,1) \}=0.
	\end{equation}
	Then we take the linear combination 
	\begin{equation}
		\Sigma_{\text{even}}\equiv \frac{1}{L_y/2}\sum_{i_y\,\in\,\text{even}} X_{L_x,i_y},
	\end{equation}
	which distinguishes states of the spectrum. The symmetry operators satisfy
	\begin{equation}
		\{	\Sigma_{\text{even}}, \Gamma_{L_x}^{\text{even}}(0,1) \}=0~~~\text{and}~~~[\Sigma_{\text{even}}, \Gamma_{L_x}^{\text{odd}}(0,1)]=0.
	\end{equation}
	The nontrivial commutation between $\Sigma_{\text{even}}$ and $\Gamma_{L_x}^{\text{even}}(0,1)$ implies that the ground state is two-fold degenerate, leading to a spontaneous symmetry breaking. The symmetry generated by $ \Gamma_{L_x}^{\text{odd}}(0,1)$ is exact.
	
	The operator $\Gamma_{L_x}^{\text{even}}(0,1)$ acts nontrivially on the ground states 
	\begin{equation}
		\bigotimes_{i_y\,\in\,\text{odd}}
		\ket{ \begin{smallmatrix}
				+
				\\
				~~\uparrow_{i_y}
				\\
				+
		\end{smallmatrix}}
		~~~\text{and}~~~
		\bigotimes_{i_y\,\in\,\text{odd}}
		\ket{ \begin{smallmatrix}
				-
				\\
				~~\uparrow_{i_y}
				\\
				-
		\end{smallmatrix}},
		\label{gs123}
	\end{equation}
	while the operator $\Sigma_{\text{even}}$ distinguishes them. Thus, these ground states spontaneously break the symmetry generated by $\Gamma_{L_x}^{\text{even}}(0,1)$ and  realize exactly the symmetry generated by  $\Sigma_{\text{even}}$. If we take linear combinations of the ground states in \eqref{gs123}, the realizations of these symmetries is interchanged. The symmetry generated by $\Gamma_{L_x}^{\text{odd}}(0,1)$ is always exactly realized.

	
	\subsubsection{Condensation of $\{_{o_1}^{e_1},_{e_2}^{o_2}\}$ }
	
	Now we discuss a boundary phase that condenses composite excitations in the group \eqref{comp}. In this case we have to consider a boundary Hamiltonian constructed from composition of the basic boundary operators \eqref{stab vertical}. A composite excitation in the lattice involves a characteristic length, namely, the distance between its constituents. We shall discuss here composite excitations separated by the minimum length (unit lattice spacing) allowed by the lattice.
	
	We first consider the boundary phase described by 
	\begin{eqnarray}
		H^v(_{o_1}^{e_1},_{e_2}^{o_2}) &=& H_{\text{bulk}}\,  - \sum_{i_y\,\in \,\text{even}}( B_{1,i_y}^v B_{1,i_y-1}^v + B_{2,i_y}^v B_{2,i_y+1}^v)\nonumber\\
		&=& 
		H_{\text{bulk}} \, - \sum_{i_y\,\in\, \text{even}} \begin{matrix}
			{}&X
			\\
			Z&Y_{i_y}
			\\
			Z & Y 
			\\
			{}&X
		\end{matrix}~
		- \sum_{i_y\,\in\, \text{even}} \begin{matrix}
			Z
			\\
			Z_{i_y}
		\end{matrix}.
		\label{111}
	\end{eqnarray}
	
	According to the Fig. \ref{trio}, we see that the composite excitations $_{o_1}^{e_1}$ and $_{e_2}^{o_2}$ are condensed at the boundary. This notation corresponds to the vertical disposition of the composite excitations. This specification is relevant since, despite the fact that $_{o_1}^{e_1}$ is condensed at the boundary, its up side down counterpart $_{e_1}^{o_1}$ is not condensed with the lattice Hamiltonian \eqref{111}.  Nevertheless, we can easily construct another boundary Hamiltonian leading to the condensation of $_{e_1}^{o_1}$. It has the same form as \eqref{111} but runs over the odd sublattice.

	Integrating out the bulk in \eqref{111}, we obtain the effective boundary Hamiltonian
	\begin{equation}
		H_{\text{eff}}^v(_{o_1}^{e_1},_{e_2}^{o_2}) =  - \sum_{i_y\,\in\, \text{even}} \begin{matrix}
			X
			\\
			Y_{i_y}
			\\
			Y 
			\\
			X
		\end{matrix}~
		- \sum_{i_y\,\in\, \text{even}} \begin{matrix}
			Z
			\\
			Z_{i_y}
		\end{matrix}.
		\label{111a}
	\end{equation}
	The set of symmetry operators is given by the extended ones 
	\begin{equation}
		\Gamma_{L_x}^{\text{even}}(0,1)~~~\text{and}~~~ \Gamma_{L_x}^{\text{odd}}(0,1)
	\end{equation}
	and the local operators
	\begin{equation}
		\begin{matrix}
			\!\!\!\!\!Y
			\\
			~~~~~~X_{L_x, i_y\,\in\, \text{even}}
		\end{matrix}
		~~~\text{and}~~~
		\begin{matrix}
			\!\!\!\!\!X
			\\
			~~~~~~Y_{L_x, i_y\,\in\, \text{even}}
		\end{matrix}.
	\end{equation}
	These two local operators are not independent as they are connected by one operator of the boundary Hamiltonian \eqref{111a}, so that it suffices to consider only one of them. We take the linear combination
	\begin{equation}
		\Phi_{\text{even}}\equiv  \frac{1}{L_y/2}\sum_{i_y\,\in \,\text{even}} \begin{matrix}
			\!\!\!	Y
			\\
			~~X_{L_x, i_y}.
		\end{matrix}
	\end{equation}
	The symmetry operators satisfy
	\begin{equation}
		\{\Phi_{\text{even}},  \Gamma_{L_x}^{\text{even}}(0,1)\}=\{\Phi_{\text{even}},  \Gamma_{L_x}^{\text{odd}}(0,1)\}=0
	\end{equation}	
	and
	\begin{equation}	
		[\Phi_{\text{even}},  \Gamma_{L_x}^{\text{even}}(0,1) \Gamma_{L_x}^{\text{odd}}(0,1)]=0.
	\end{equation}
	These relations imply that the ground state is two-fold degenerate, corresponding to a spontaneous symmetry-breaking phase. The diagonal symmetry is $ \Gamma_{L_x}^{\text{even}}(0,1) \Gamma_{L_x}^{\text{odd}}(0,1)$ is exact.

	We can construct the degenerate ground states of \eqref{111a} as 
	\begin{equation}
		\prod_{i_y\,\in\,\text{even}} A_{i_y}  	\bigotimes_{i_y} \ket{\uparrow_{i_y}}~~~\text{and}~~~ \Phi_{\text{even}}\left(  \prod_{i_y\,\in\,\text{even}} A_{i_y}  	\bigotimes_{i_y} \ket{\uparrow_{i_y}} \right),
		\label{sdfg}
	\end{equation}
	where $A_{i_y}$ is a projector defined as
	\begin{equation}
		A_{i_y}\equiv \frac12 \left( \openone~ +\begin{smallmatrix}
			X
			\\
			~Y_{i_y}
			\\
			Y 
			\\
			X
		\end{smallmatrix} \right).
	\end{equation}
	The symmetry operators $ \Gamma_{L_x}^{\text{even}}(0,1)$ and $\Gamma_{L_x}^{\text{odd}}(0,1)$ act trivially on the ground states in \eqref{sdfg}, whereas the operator $\Phi_{\text{even}}$ connects them. This implies that such ground states spontaneously break the symmetries generated by $\Phi_{\text{even}}$, while the symmetries generated by $ \Gamma_{L_x}^{\text{even}}(0,1)$ and  $\Gamma_{L_x}^{\text{odd}}(0,1)$ are exact. Taking linear combinations of the ground states in \eqref{sdfg} interchanges these realizations of symmetries. The diagonal symmetry generated by $ \Gamma_{L_x}^{\text{even}}(0,1) \Gamma_{L_x}^{\text{odd}}(0,1)$, however, is always exactly realized.

	
	\subsubsection{Condensation of $\{e_1 e_2,o_1 o_2\}$ }
	
	Finally, we discuss the boundary phase that condenses the composite excitations $e_1 e_2$ and $o_1 o_2$. The corresponding Hammiltonian reads
	\begin{eqnarray}
		H^v(e_1 e_2,o_1 o_2) &=&H_{\text{bulk}} \,- \sum_{i_y\,\in\, \text{even}} B_{1,i_y}^v B_{2,i_y}^v -  \sum_{i_y\,\in\, \text{odd}} B_{1,i_y}^v B_{2,i_y}^v \nonumber\\
		&=&H_{\text{bulk}} \, - \sum_{i_y} \,
		\begin{matrix}
			{}&X
			\\
			Z&\openone_{i_y}
			\\
			{} & X 
		\end{matrix}.
	\end{eqnarray}
	Integrating out the bulk, the boundary effective theory reduces to
	\begin{equation}
		H_{\text{eff}}^v(e_1 e_2,o_1 o_2) = - \sum_{i_y} \,
		\begin{matrix}
			X
			\\
			\openone_{i_y}
			\\
			X 
		\end{matrix}.
		\label{tyu}
	\end{equation}
	
	In addition to the extended symmetry operators 
	\begin{equation}
		\Gamma_{L_x}^{\text{even}}(0,1)~~~\text{and}~~~ \Gamma_{L_x}^{\text{odd}}(0,1),
	\end{equation}
	we have the local symmetry operators
	\begin{equation}
		X_{L_x, i_y\,\in\,\text{even}}~~~\text{and}~~~	X_{L_x, i_y\,\in\,\text{odd}},
	\end{equation}
	from which we can construct
	\begin{equation}
		\Sigma_{\text{even}}\equiv \frac{1}{L_y/2}\sum_{i_y\,\in\,\text{even}} X_{L_x,i_y}~~~\text{and}~~~	
		\Sigma_{\text{odd}}\equiv \frac{1}{L_y/2}\sum_{i_y\,\in\,\text{odd}} X_{L_x,i_y}.
	\end{equation}
	
	The set of symmetry operators satisfies
	\begin{equation}
		\{	\Sigma_{\text{even}}, \Gamma_{L_x}^{\text{even}}(0,1) \}=0~~~\text{and}~~~ 	\{	\Sigma_{\text{odd}}, \Gamma_{L_x}^{\text{odd}}(0,1) \}=0,
	\end{equation}
	and
	\begin{equation}
		[	\Sigma_{\text{even}}, \Gamma_{L_x}^{\text{odd}}(0,1) ]=0~~~\text{and}~~~ [\Sigma_{\text{odd}}, \Gamma_{L_x}^{\text{even}}(0,1)]=0.
	\end{equation}
	These relations imply that the ground state is four-fold degenerate, corresponding to a spontaneous symmetry-breaking phase.
	
	From \eqref{tyu}, we construct immediately the ground states
	\begin{eqnarray}
		\ket{++}\equiv\bigotimes_{i_y\,\in\,\text{even}} \ket{+_{i_y}} &\otimes&  \bigotimes_{i_y\,\in\,\text{odd}} \ket{+_{i_y}};  \nonumber\\
		\ket{+-}\equiv\bigotimes_{i_y\,\in\,\text{even}} \ket{+_{i_y}} &\otimes&  \bigotimes_{i_y\,\in\,\text{odd}} \ket{-_{i_y}};  \nonumber\\
		\ket{-+}\equiv\bigotimes_{i_y\,\in\,\text{even}} \ket{-_{i_y}} &\otimes&  \bigotimes_{i_y\,\in\,\text{odd}} \ket{+_{i_y}};  \nonumber\\
		\ket{--}\equiv\bigotimes_{i_y\,\in\,\text{even}} \ket{-_{i_y}} &\otimes&  \bigotimes_{i_y\,\in\,\text{odd}} \ket{-_{i_y}}.
		\label{xcv}
	\end{eqnarray}
	The symmetry operators $ \Gamma_{L_x}^{\text{even}}(0,1)$ and $\Gamma_{L_x}^{\text{odd}}(0,1)$, and their product $\Gamma_{L_x}^{\text{even}}(0,1)\Gamma_{L_x}^{\text{odd}}(0,1)$, connect all the above  ground states, while the operators $	\Sigma_{\text{even}}$ and $	\Sigma_{\text{odd}}$ act trivially on them. Thus, the ground states \eqref{xcv} spontaneously break the symmetries generated by $ \Gamma_{L_x}^{\text{even}}(0,1)$ and $\Gamma_{L_x}^{\text{odd}}(0,1)$, while the symmetries generated by  $	\Sigma_{\text{even}}$ and $	\Sigma_{\text{odd}}$ are exact. 
	
	Orthogonal linear combinations of the ground states in \eqref{xcv},
	\begin{eqnarray}
		&&\ket{++}\,+\,\ket{+-}\,+\,\ket{-+}\,+\,\ket{--};\nonumber\\
		&&\ket{++}\,-\,\ket{+-}\,+\,\ket{-+}\,-\,\ket{--};\nonumber\\
		&&\ket{++}\,+\,\ket{+-}\,-\,\ket{-+}\,-\,\ket{--};\nonumber\\
		&&\ket{++}\,-\,\ket{+-}\,-\,\ket{-+}\,+\,\ket{--},
		\label{xcv1}
	\end{eqnarray}
	interchange the realization of the symmetries, i.e., $ \Gamma_{L_x}^{\text{even}}(0,1)$ and $\Gamma_{L_x}^{\text{odd}}(0,1)$ act trivially on them, while the operators 
	$\Sigma_{\text{even}}$ and $\Sigma_{\text{odd}}$, and their product $\Sigma_{\text{even}}\Sigma_{\text{odd}}$, connect the ground states in \eqref{xcv1}. Thus, the states \eqref{xcv1} spontaneously break the  symmetries generated by $\Sigma_{\text{even}}$ and $\Sigma_{\text{odd}}$, whereas the symmetries generated by $ \Gamma_{L_x}^{\text{even}}(0,1)$ and $\Gamma_{L_x}^{\text{odd}}(0,1)$ are exactly realized.

	
	\section{Effective Field Theory: Low Energies and Short Distances}\label{eft01}

	So far we have discussed the physical properties from the lattice perspective. The topological phase captured by the lattice Hamiltonian exhibits UV/IR mixing that manifests, for example, in the fact that the number of degenerate ground states is affected by the number of sites. This type of feature is shared with many other lattice models \cite{Bulmash:2018knk,Oh_2022,Oh:2023bnk,Pace_2022,Casasola:2023tot,Casasola2024,Watanabe:2022pgk} and also with topologically ordered fractonic systems, which exhibit a much stronger form of UV/IR mixing.

	The UV/IR mixing poses an important question concerning the very existence of such phases, as it challenges effective low-energy and continuum descriptions. We address this type of question in a two-steps process. First, we derive an effective description in low energies but not long distances. In other words, we describe effectively the low-energy excitations but keep the lattice details in the analysis. This approach makes transparent the role of the generalized modulated symmetries in the mobility of excitations and incorporates the boundary physics in a nice way. In a second step, taken in the next section, we consider a low-energy and long-distance description capturing some universal data of the phase.

	\subsection{Embedding in $U(1)$ Theory}
	
	Considering that $L_y$ is even, we split the square lattice in odd and even sublattices along the $y$-direction and attach complex bosonic degrees of freedom $\Phi_{\vec{r}}$ to one of them, say to the odd sublattice, specified by $\vec{r}=(i_x,i_y)$, with $i_y\,\in\,\text{odd}$. Degrees of freedom in the even sites are introduced posteriorly along with the gauging procedure. Then, we set down a Hamiltonian capturing the local dynamics of the excitations,
	\begin{equation}
		H= \frac{1}{2}\sum_{\vec{r} \,\in\,\text{odd}} \Pi_{\vec{r}}^{\dagger} \Pi_{\vec{r}} -t_{\text{h}} \sum_{\vec{r}\,\in\,\text{odd}}\Phi_{\vec{r}-\hat{x}}\Phi_{\vec{r}} \Phi_{\vec{r}+\hat{x}} -
		t_{\text{v}} \sum_{\vec{r}\,\in\,\text{odd}}\Phi_{\vec{r}}^{\dagger}\Phi_{\vec{r}+2\hat{y}}+ \text{H. c.} + \cdots,
		\label{vvv}
	\end{equation} 
	where $\Pi_{\vec{r}}$ is the canonical momentum conjugated to the bosonic field $\Phi_{\vec{r}}$, and the dots account for a Higgs potential. The notation $\vec{r} \,\in\,\text{odd}$ in the summations means that the vertical sums run over $i_y\,\in\,\text{odd}$.
	
	The Hamiltonian is invariant under the $U(1)$ nonuniform global symmetries 
	\begin{equation}
		\Phi_{\vec{r}} \rightarrow e^{i f_{\vec{r}}} \Phi_{\vec{r}},
	\end{equation}
	where the function $f_{\vec{r}}$ satisfies
	\begin{equation}
		f_{\vec{r}-\hat{x}} + f_{\vec{r}}+f_{\vec{r}+\hat{x}}=0~~~\text{and}~~~ f_{\vec{r}-\hat{y}} -f_{\vec{r}+\hat{y}}=0,
		\label{cvb}
	\end{equation}
	which are nothing else the $U(1)$ version of the constraints \eqref{eq:const1} and \eqref{eq:const2}. As such, the existence of nontrivial solutions depends on the linear size $L_x$ of the system. If $L_x$ is not a multiple of three, \eqref{cvb} has only the trivial solution $f_{\vec{r}}=0$.  If $L_x$ is a multiple of three we have two linearly independent solutions. Let us write the solutions of \eqref{cvb} as
	\begin{equation}
		f_{\vec{r}} = \theta_{\mu} f_{i_x}^{(\mu)},~~~\mu=1,2,
		\label{cvb1}
	\end{equation}
	where $\theta_{\mu}$ are the $U(1)$ global parameters and $ f_{i_x}^{(\mu)}$ are an integer-valued functions satisfying 
	\begin{equation}
		f_{i_x-1}^{(\mu)}+f_{i_x}^{(\mu)}+f_{i_x+1}^{(\mu)}=0,
		\label{jkl}
	\end{equation}
	with values restricted to $+1$, $-1$, and $0$. We choose the two linearly independent solutions as 
	\begin{equation}
		f_{i_x}^{(1)}=(-1,1,0,-1,1,0\ldots,-1,1,0)~~~\text{and}~~~f_{i_x}^{(2)}=(0,1,-1,0,1,-1\ldots,0,1,-1).
		\label{qqq}
	\end{equation}
	If $L_x$ is not a multiple of three, $f_{i_x}^{(\mu)}=0$.

	The next step is to gauge such global symmetries. It is more convenient to convert to the Lagrangian formalism 
	\begin{equation}
		S= \int dt\, \ell_x \ell_y \sum_{\vec{r}\,\in\,\text{odd}} \left[ |\partial_t\Phi_{\vec{r}}|^2 + t_{\text{h}}\, \Phi_{\vec{r}-\hat{x}}\Phi_{\vec{r}} \Phi_{\vec{r}+\hat{x}} +
		t_{\text{v}}\,\Phi_{\vec{r}}^{\dagger}\Phi_{\vec{r}+2\hat{y}}+ \text{H. c.} +\cdots\right],
	\end{equation}
	where $\ell_x$ and $\ell_y$ are the lattice spacings with length units. The gauged version of this action is simply
	\begin{eqnarray}
		S &=& \int dt\, \ell_x \ell_y  \sum_{\vec{r}\,\in\,\text{odd}} \left[ \mathcal{L}(a^t_{\vec{r}}, a^x_{\vec{r}}, a^y_{\vec{r}+\hat{y}})+ |(\partial_t-i q a_{\vec{r}}^t)\Phi_{\vec{r}}|^2 \right. \nonumber \\
		&+& \left. t_{\text{h}}\, e^{-i q \ell_x a^x_{\vec{r}}} \Phi_{\vec{r}-\hat{x}}\Phi_{\vec{r}} \Phi_{\vec{r}+\hat{x}} +
		t_{\text{v}} \, e^{-i q \ell_y a^y_{\vec{r}+\hat{y}}}\Phi_{\vec{r}}^{\dagger}\Phi_{\vec{r}+2\hat{y}}+ \text{H. c.}+\cdots\right],
		\label{bnm}
	\end{eqnarray}
	where we are considering that $\Phi_{\vec{r}}$ has gauge integer-charge $q$. The case $q=2$ captures the physics of the lattice model. The dynamics of the gauge fields is governed by a Maxwell term  
	\begin{equation}
		\mathcal{L}(a^t_{\vec{r}}, a^x_{\vec{r}}, a^y_{\vec{r}}) =\frac{1}{2g_x}(\Delta_x a^t_{\vec{r}}-\partial_ta^x_{\vec{r}})^2 +\frac{1}{2g_y}(\Delta_y a^t_{\vec{r}}-\partial_t a^y_{\vec{r}+\hat{y}})^2- \frac{1}{2g}(\Delta_x a^y_{\vec{r}+\hat{y}}-\Delta_y a^x_{\vec{r}})^2,
	\end{equation}
	and the operators $\Delta_x$ and $\Delta_y$ are defined as
	\begin{equation}
		\Delta_x\Lambda_{\vec{r}}\equiv \frac{1}{\ell_x}(\Lambda_{\vec{r}-\hat{x}}+\Lambda_{\vec{r}}+\Lambda_{\vec{r}+\hat{x}})~~~\text{and}~~~
		\Delta_y\Lambda_{\vec{r}}\equiv \frac{1}{\ell_y}(\Lambda_{\vec{r}+2\hat{y}}-\Lambda_{\vec{r}}).
	\end{equation}
	
	The action \eqref{bnm} is invariant under the gauge transformations
	\begin{eqnarray}
		a^t_{\vec{r}}&\rightarrow& a^t_{\vec{r}} + \partial_t\Lambda_{\vec{r}}; \nonumber\\
		a^x_{\vec{r}}&\rightarrow& a^x_{\vec{r}} + \Delta_x\Lambda_{\vec{r}}; \nonumber\\
		a^y_{\vec{r}+\hat{y}}&\rightarrow& a^y_{\vec{r}+\hat{y}} + \Delta_y\Lambda_{\vec{r}};\nonumber\\
		\Phi_{\vec{r}} &\rightarrow& e^{i q \Lambda_{\vec{r}}} \Phi_{\vec{r}},
		\label{gt}
	\end{eqnarray}
	with $\Lambda_{\vec{r}}$ being an arbitrary gauge function.

	Next we go to the Higgs phase. Its deep IR can be parametrized as
	\begin{equation}
		\Phi_{\vec{r}} \rightarrow \rho_0 e^{i \phi_{\vec{r}}},
	\end{equation}
	where $\rho_0$ is the frozen radial mode, with $\rho_0\rightarrow \infty$. The corresponding action reduces to 
	\begin{eqnarray}
		S_{\text{Higgs}} &=& \int dt\, \ell_x \ell_y  \sum_{\vec{r}\,\in\,\text{odd}} \left[ \mathcal{L}(a^t_{\vec{r}}, a^x_{\vec{r}}, a^y_{\vec{r}+\hat{y}})+ \rho_0^2(\partial_t\phi_{\vec{r}}- q a_{\vec{r}}^t)^2 \right. \nonumber \\
		&+& \left. 2 \rho_0^3t_{\text{h}}\, \cos[\ell_x(\Delta_x\phi_{\vec{r}}- q a^x_{\vec{r}})] +
		2 \rho_0^2 t_{\text{v}} \,  \cos[\ell_y(\Delta_y\phi_{\vec{r}}- q a^y_{\vec{r}+\hat{y}})] \right].
		\label{bnmH}
	\end{eqnarray}
	We proceed to express the theory only in terms of gauge fields. For this, we introduce new gauge fields as
	\begin{equation}
		\partial_t\phi_{\vec{r}} \equiv c^t_{\vec{r}},~~~ \Delta_x\phi_{\vec{r}}\equiv c^x_{\vec{r}},~~~  \Delta_y\phi_{\vec{r}}\equiv c^y_{\vec{r}+\hat{y}},
	\end{equation}
	satisfying the constraints of ``zero flux",
	\begin{equation}
		\Delta_x c^y_{\vec{r}+\hat{y}}-\Delta_y c^x_{\vec{r}} = 0,~~~ \Delta_y c^t_{\vec{r}}-\partial_t c^y_{\vec{r}+\hat{y}} = 0,~~~ \Delta_x c^t_{\vec{r}}-\partial_t c^x_{\vec{r}}=0. 
	\end{equation}
	These constraints can be taken into account in the action through Lagrange multipliers $b^t$, $b^x$, and $b^y$. We have
	\begin{eqnarray}
		S_{\text{Higgs}} &=& \int dt\, \ell_x \ell_y  \sum_{\vec{r}\,\in\,\text{odd}} \left[ \mathcal{L}(a^t_{\vec{r}}, a^x_{\vec{r}}, a^y_{\vec{r}+\hat{y}})+ \rho_0^2(c^t_{\vec{r}}- q a_{\vec{r}}^t)^2 \right. \nonumber \\
		&+&2 \rho_0^3t_{\text{h}}\, \cos[\ell_x(c^x_{\vec{r}}- q a^x_{\vec{r}})] +
		2 \rho_0^2 t_{\text{v}} \,  \cos[\ell_y(c^y_{\vec{r}+\hat{y}}- q a^y_{\vec{r}})] \nonumber\\
		&+& \left. \frac{1}{2\pi} \big(b^t_{\vec{r}+\hat{y}}(\Delta_x c^y_{\vec{r}+\hat{y}}-\Delta_y c^x_{\vec{r}})
		+b^x_{\vec{r}+\hat{y}}(\Delta_y c^t_{\vec{r}}-\partial_t c^y_{\vec{r}})
		- b^y_{\vec{r}}(\Delta_x c^t_{\vec{r}}-\partial_t c^y_{\vec{r}+\hat{y}})\big) \right].
		\label{bnmDual}
	\end{eqnarray}
	In the deep IR limit, $\rho_0\rightarrow \infty$, the integration over the $c$-fields leads to 
	\begin{equation}
		S_{\text{Higgs}} =  \int dt\, \ell_x \ell_y  \sum_{\vec{r}\,\in\,\text{odd}} \frac{q}{2\pi} \big[b^t_{\vec{r}+\hat{y}}(\Delta_x a^y_{\vec{r}+\hat{y}}-\Delta_y a^x_{\vec{r}})
		+b^x_{\vec{r}+\hat{y}}(\Delta_y a^t_{\vec{r}}-\partial_t a^y_{\vec{r}+\hat{y}})
		- b^y_{\vec{r}}(\Delta_x a^t_{\vec{r}}-\partial_t a^x_{\vec{r}})\big],
		\label{bf}
	\end{equation}
	which is a BF-like action involving $\mathbb{Z}_q$ symmetries. The lattice model corresponds to the $\mathbb{Z}_2$ case. 
	In addition to the gauge transformations of the $a$-fields given in \eqref{gt}, the action \eqref{bf} is also invariant under gauge transformations of the $b$-fields,
	\begin{eqnarray}
		b^t_{\vec{r}+\hat{y}}&\rightarrow& b^t_{\vec{r}+\hat{y}} + \partial_t\Omega_{\vec{r}+\hat{y}}; \nonumber\\
		b^x_{\vec{r}+\hat{y}}&\rightarrow& b^x_{\vec{r}+\hat{y}} - \Delta_x\Omega_{\vec{r}+\hat{y}}; \nonumber\\
		b^y_{\vec{r}}&\rightarrow& b^y_{\vec{r}} + \Delta_y\Omega_{\vec{r}+\hat{y}},
	\end{eqnarray}
	where $\Omega_{\vec{r}+\hat{y}}$ is an arbitrary gauge function.

		It is interesting to contrast the theory \eqref{bf} with the BF-like theories involved in description of other  somewhat similar systems, as discussed in Refs. 
		\cite{Bulmash_2018,Oh_2022,Pace_2022,Delfino_2023,Watanabe:2022pgk,Oh:2023bnk,PhysRevB.106.155150,Ebisu:2023idd,PhysRevB.109.235127}. 
		In these cases, the BF-like descriptions are associated with multipolar symmetries. For example, in the dipolar case, the BF-like description incorporates both charge and dipole conservation. As a consequence, typically such theories involve higher-rank gauge fields.

	Now we can extract physical information from the effective field theory \eqref{bf}.

	
	\subsection{Defects and Generalized Modulated Symmetries}
	
	We will analyze first the defects, which are lines extended along time direction
	\begin{equation}
		\exp\left(i\, m_t \oint dt\, a^t_{\vec{r}}\right)~~~\text{and}~~~\exp\left(i\, n_t \oint dt\, b^t_{\vec{r}+\hat{y}}\right).
		\label{eee}
	\end{equation}
	The physical interpretation of these objects is that they describe single excitations at rest. As we shall see below, the parameters $m_t$ and $n_t$ are integers due to large gauge transformations
	\begin{equation}
		\Lambda_{\vec{r}} = \Omega_{\vec{r}+\hat{y}}= 2\pi \frac{t}{L_t} p,~~ p \in \mathbb{Z},
		\label{fff}
	\end{equation}
	where $L_t$ is the dimensional length of the time direction, which is assumed to be a circle.  
	
	In what follows, we shall discuss only the transformations of the $a$-fields, as the transformations of the $b$-fields are identical.
	
	Under a gauge transformation with the gauge function \eqref{fff}, the $a$-fields transform as
	\begin{equation}
		a^t_{\vec{r}}\rightarrow a^t_{\vec{r}} +   \frac{2\pi}{L_t} p,~~~ a^x_{\vec{r}} \rightarrow  a^x_{\vec{r}} +  6\pi \frac{t}{\ell_x L_t} n ,~~~  a^y_{\vec{r}} \rightarrow  a^y_{\vec{r}}.
	\end{equation}
	The requirement that the defect involving the field $a_{\vec{r}}^t$ in \eqref{eee} is invariant under such transformation implies that $m_t$ must be an integer.
	
	Another type of large gauge transformations are the nonuniform ones, 
	\begin{equation}
		\Lambda_{\vec{r}} = 2\pi \frac{t}{L_t} f_{i_x}^{(\mu)},
	\end{equation}
	with $f_{i_x}^{(\mu)}$ being the integer-valued functions introduced in \eqref{qqq}. This leads to the following transformations of the gauge fields
	\begin{equation}
		a^t_{\vec{r}}\rightarrow a^t_{\vec{r}} +   \frac{2\pi}{L_t}  f_{i_x}^{(\mu)},~~~ a^x_{\vec{r}} \rightarrow  a^x_{\vec{r}},~~~  a^y_{\vec{r}+\hat{y}} \rightarrow  a^y_{\vec{r}+\hat{y}},
		\label{lll}
	\end{equation}
	which also leave invariant the defect in \eqref{eee}.
	
	In contrast with \eqref{lll}, we can consider the transformations
	\begin{equation}
		a^t_{\vec{r}}\rightarrow a^t_{\vec{r}} +  \frac{2\pi}{q} \frac{f_{i_x}^{(\mu)}}{L_t},~~~ a^x_{\vec{r}} \rightarrow  a^x_{\vec{r}},~~~  a^y_{\vec{r}+\hat{y}} \rightarrow  a^y_{\vec{r}+\hat{y}}.
		\label{lll1}
	\end{equation}
	Although they look like the gauge transformations in \eqref{lll}, the changing of $a^t_{\vec{r}}$ differs in an essential way by a factor of $1/q$, which makes \eqref{lll1} a genuine global symmetry of the action \eqref{bf} that cannot be undone by any gauge transformation. Furthermore, the defects in \eqref{eee} are charged under such symmetry, 
	\begin{equation}
		\exp\left(i\, m_t \oint dt\, a^t_{\vec{r}}\right) ~\rightarrow~  \exp\left(  \frac{2\pi}{q} f_{i_x}^{(\mu)}\right) \exp\left(i\, m_t \oint dt\, a^t_{\vec{r}}\right),
	\end{equation}
	with the factor $\exp\left(  \frac{2\pi}{q} f_{i_x}^{(\mu)}\right)$ corresponding to the charge of the defect under the global symmetry. As the charge depends on the position $i_x$, its conservation implies that the excitation can only move in steps of size three, which is the periodicity of the function $ f_{i_x}^{(\mu)}$. Otherwise, the global symmetry would be violated. Also, as $i_{y}\,\in\,\text{odd}$, the excitation is restricted to the odd sublattice. This is how the mobility restriction of excitations can be understood as a direct consequence of the generalized modulated symmetries.

	
	\subsection{Symmetry Operators and 't Hooft Anomalies}

	Now we discuss the symmetry operators, which are objects extended along spatial directions at a fixed time. 
	
	Symmetry operators along the $x$-direction are
	\begin{equation}
		\exp\left( i\, m^{x}_{\mu}\, \ell_x\sum_{i_x}  f_{i_x}^{(\mu)} a_{\vec{r}}^x\right)~~~\text{and}~~~\exp\left( i\, n^{x}_{\mu}\, \ell_x\sum_{i_x}   f_{i_x}^{(\mu)} b_{\vec{r}+\hat{y}}^x\right),
		\label{ert}
	\end{equation}
	which are modulated by the functions $ f_{i_x}^{(\mu)} $. 
	The factors $m_{\mu}^x$ and $n_{\mu}^x$ are also integers due to large gauge transformations. A large gauge transformation that winds around the $x$-direction is 
	\begin{equation}
		\Lambda_{\vec{r}} = 2\pi \frac{i_x}{L_x} p,~~~p \in \mathbb{Z},
	\end{equation}
	where $L_x$ is the dimensionless length. Under this transformation, the $a$-fields transform as
	\begin{equation}
		a^t_{\vec{r}}\rightarrow a^t_{\vec{r}},~~~ a^x_{\vec{r}} \rightarrow  a^x_{\vec{r}} +  6\pi \frac{i_x}{\ell_x L_x}p  ,~~~  a^y_{\vec{r}+\hat{y}} \rightarrow  a^y_{\vec{r}+\hat{y}}.
	\end{equation}
	The symmetry operator involving the $a^x$-field in \eqref{ert} changes by a phase factor
	\begin{equation}
		\exp\left(i \,m_{\mu}^x  \frac{6\pi}{L_x} \sum_{i_x} f_{i_x}^{(\mu)}i_x \right).
	\end{equation} 
	To evaluate the sum over $i_x$, we write it as 
	\begin{equation}
		\sum_{i_x}  f_{i_x}^{(\mu)} i_x= f_{i_x=1}^{(\mu)}(1+4+7+\cdots) +  f_{i_x=2}^{(\mu)}(2+5+8+\cdots)+ f_{i_x=3}^{(\mu)}(3+6+9+\cdots).
	\end{equation}
	With the choices for $f_{i_x}^{(1)}$ and $f_{i_x}^{(2)}$ in \eqref{qqq}, the outcomes are
	\begin{equation}
		\sum_{i_x} f_{i_x}^{(1)} i_x=  \frac{L_x}{3}~~~\text{and}~~~\sum_{i_x} f_{i_x}^{(2)} i_x=  - \frac{L_x}{3}.
	\end{equation}
	The requirement that the symmetry operator is invariant under such a large gauge transformation implies that the parameters $m^x_{\mu}$ must be integers.

	The symmetry operators along the $y$-direction are
	\begin{equation}
		\exp\left( i\, m^{y}\, \ell_y\sum_{i_y\,\in\,\text{odd}}  a_{\vec{r}+\hat{y}}^y\right)~~~\text{and}~~~\exp\left( i\, n^{y}\, \ell_y\sum_{i_y\,\in\,\text{odd}}   b_{\vec{r}}^y\right).
		\label{ert2}
	\end{equation}
	The large gauge transformation
	\begin{equation}
		\Lambda_{\vec{r}} = 2\pi \frac{i_y}{L_y} p,~~~p \in \mathbb{Z},
	\end{equation}
	induces 
	\begin{equation}
		a^t_{\vec{r}}\rightarrow a^t_{\vec{r}},~~~ a^x_{\vec{r}} \rightarrow  a^x_{\vec{r}} +  6\pi \frac{i_y}{\ell_x L_y}p  ,~~~  a^y_{\vec{r}+\hat{y}} \rightarrow  a^y_{\vec{r}+\hat{y}}- \frac{4\pi}{\ell_y L_y}p.
	\end{equation}
	Invariance under these transformations implies that $m^y$ must be an integer.

	After the construction of the symmetry operators, we are ready to analyze the 't Hooft anomalies between them. First, we need the commutation relations between the gauge fields, which follow from \eqref{bf}, 
	\begin{equation}
		[a^x_{\vec{r}}, b_{\vec{r}'}^y]=\frac{2\pi i}{q} \frac{ \delta_{\vec{r},\vec{r}'}}{\ell_x\ell_y}~~~\text{and}~~~ [a^y_{\vec{r}+\hat{y}}, b_{\vec{r}'+\hat{y}}^x]=-\frac{2\pi i}{q} \frac{ \delta_{\vec{r},\vec{r}'}}{\ell_x\ell_y}.
		\label{cr12}
	\end{equation}
	We emphasize that the fields $a^x_{\vec{r}}$ and $ b_{\vec{r}'}^y$ live at the odd sublattice, while $a^y_{\vec{r}+\hat{y}}$ and $ b_{\vec{r}'+\hat{y}}^x$ live at the even one.
	
	For the operators of the odd sublattice, it follows that
	\begin{equation}
		\exp\left( i \ell_x\sum_{i_x}  f_{i_x}^{(\mu)} a_{\vec{r}}^x\right) \exp\left( i \ell_y\sum_{j_y\,\in\,\text{odd}}   b_{\vec{r}'}^y\right)= e^{-\frac{2\pi i}{q}f_{j_x}^{(\mu)}}
		\exp\left( i \ell_y\sum_{j_y\,\in\,\text{odd}}   b_{\vec{r}'}^y\right)\exp\left( i \ell_x\sum_{i_x}  f_{i_x} a_{\vec{r}}^x\right), 
		\label{erty}
	\end{equation}
	with $\vec{r}=(i_x,i_y)$ and $\vec{r}'=(j_x,j_y)\,\in\, \text{odd}$. Let us choose the independent vertical operators as the ones localized at $j_x= 1 \!\!   \mod 3$ and  $j_x= 3 \!\!   \mod 3$. Then, the horizontal operators specified by $f^{(1)}_{j_x =1\!\!   \mod 3}=-1$ and $f^{(2)}_{j_x =3\!\!   \mod 3}=-1$ will have nontrivial commutation relations with the vertical operators, corresponding to two 't Hooft anomalies.

	Similarly, for the operators of the even sublattice, we have 
	\begin{equation}
		\exp\left( i \ell_x\sum_{i_x}   f_{i_x}^{(\mu)} b_{\vec{r}}^x\right) \exp\left( i\ell_y\sum_{i_y\,\in\,\text{even}}  a_{\vec{r}'}^y\right)=e^{-\frac{2\pi i}{q}f_{j_x}^{(\mu)}}
		\exp\left( i\ell_y\sum_{i_y\,\in\,\text{even}}  a_{\vec{r}'}^y\right)\exp\left( i \ell_x\sum_{i_x}   f_{i_x}^{(\mu)} b_{\vec{r}}^x\right),
		\label{erty1}
	\end{equation}
	which leads to two more independent  't Hooft anomalies for $f^{(1)}_{j_x =1\!\!   \mod 3}=-1$ and $f^{(2)}_{j_x =3\!\!   \mod 3}=-1$. Therefore, the total number of 't Hooft anomalies is four, implying a ground state degeneracy $q^4$, which matches the lattice result for $q=2$. If $L_x$ is not a multiple of three, $f_{j_x}^{(\mu)}=0$ and there are no 't Hooft anomalies. Consequently, the ground state is unique.
	
	For $L_y$ odd, we can proceed in general lines in the same way as in the even case, but we have to pay attention to the number of degrees of freedom encoded in the resulting effective field theory. In the odd case, the starting point Hamiltonian is like the one in \eqref{vvv}, but with the sums in $\vec{r}$ running over the whole lattice, not only over the odd sites. This means that we are attaching bosonic fields to all sites of the square lattice and, in effect, duplicating the number of degrees of freedom. Therefore, we will have the same structure of symmetry operators \eqref{ert} and \eqref{ert2} and the corresponding anomalies \eqref{erty} and \eqref{erty1}, but with all sums in $i_y$ running over the whole lattice.
	This implies that we still have four 't Hooft anomalies. To compare with the lattice model, however, we need to halve the effective field theory to match the number of degrees of freedom, resulting only in two 't Hooft anomalies, so that the corresponding ground state degeneracy is $q^2$.


	\subsection{Boundary Theory}
	
	The effective field theory studied in the previous section incorporates the boundary physics in a very natural way. This follows from the fact that the action \eqref{bf} is gauge invariant only up to surface terms. Thus, in the presence of a physical boundary, gauge invariance is broken and consequently the gauge degrees of freedom at the boundary become physical modes.
	
	We discuss below the case of a vertical boundary, but the horizontal case can be treated in a similar way.

	\subsubsection{Vertical Boundary}

	Let us introduce a vertical boundary localized at $i_x = L_x$. The horizontal symmetry operators of the bulk give rise to degrees of freedom at the boundary. To see this, we consider the action \eqref{bf} and integrate out the temporal components of the gauge fields, $a_{\vec{r}}^t$ and $b_{\vec{r}+\hat{y}}^t$. This gives rise to the zero flux contraints
	\begin{equation}
		\Delta_x a^y_{\vec{r}+\hat{y}} -\Delta_y a^x_{\vec{r}}=0~~~\text{and}~~~ 	\Delta_x b^y_{\vec{r}} + \Delta_y b^x_{\vec{r}+\hat{y}}=0,
	\end{equation}
	whose solutions are pure gauge
	\begin{equation}
		a^x_{\vec{r}} = \Delta_x \varphi_{\vec{r}}~~~\text{and}~~~ a^y_{\vec{r}+\hat{y}} = \Delta_y \varphi_{\vec{r}},
		\label{par1}
	\end{equation}
	and 
	\begin{equation}
		b^x_{\vec{r}+\hat{y}} =- \Delta_x \theta_{\vec{r}+\hat{y}}~~~\text{and}~~~ b^y_{\vec{r}+\hat{y}} = \Delta_y \theta_{\vec{r}+\hat{y}}.
		\label{par2}
	\end{equation}
	The fields $\varphi_{\vec{r}}$ and $\theta_{\vec{r}+\hat{y}}$ are just gauge modes in the bulk, but they correspond to physical degrees of freedom at the boundary. 
	
	With the parametrizations in \eqref{par1} and \eqref{par2}, the horizontal symmetry operators become
	\begin{equation}
		\exp\left(i\ell_x\sum_{i_x} f_{i_x}^{(\mu)}a^x_{\vec{r}}\right)= \exp i \left(\varphi_{L_x,i_y}(f^{(\mu)}_{L_x-1}+f^{(\mu)}_{L_x})+\varphi_{L_x+1,i_y}f^{(\mu)}_{L_x}\right)
	\end{equation}
	and
	\begin{equation}
		\exp\left(i\ell_x\sum_{i_x} f_{i_x}^{(\mu)}b^x_{\vec{r}+\hat{y}}\right)= \exp i\left(-\theta_{L_x,i_y+1}(f^{(\mu)}_{L_x-1}+f^{(\mu)}_{L_x})-\theta_{L_x+1,i_y+1}f^{(\mu)}_{L_x}\right).
	\end{equation}
	Employing the functions $f_{ix}^{(\mu)}$ in \eqref{qqq}, we obtain four independent vertex operators at the boundary
	\begin{equation}
		e^{i \varphi_{L_x,i_y}},~~~ e^{-i\varphi_{L_x+1,i_y}},~~~e^{-i \theta_{L_x,i_y+1}},~~~e^{i\theta_{L_x+1,i_y+1}},
		\label{vertex}
	\end{equation}
	which involve the boundary physical fields. 
	
	The canonical commutation relations in the bulk give rise to nontrivial commutations at the boundary. From the first equation in \eqref{cr12}, we have
	\begin{equation}
		\left[\varphi_{L_x,i_y}(f^{(\mu)}_{L_x-1}+f^{(\mu)}_{L_x})+\varphi_{L_x+1,i_y}f^{(\mu)}_{L_x},\Delta_y \theta_{L_x,j_y+1}\right] = \frac{2 \pi i}{q}f^{(\mu)}_{L_x}\frac{\delta_{i_y,j_y}}{\ell_y}.
	\end{equation}
	The function $f_{i_x}^{(2)}$ gives the nontrivial relation 
	\begin{equation}
		\left[\varphi_{L_x+1,i_y},\Delta_y \theta_{L_x,j_y+1}\right] = \frac{2 \pi i}{q} \frac{\delta_{i_y,j_y}}{\ell_y}.
		\label{31}
	\end{equation}
	From the second equation in \eqref{cr12}, we have similarly
	\begin{equation}
		\left[\Delta_y\varphi_{L_x,i_y}, \theta_{L_x+1,j_y+1}\right] = -\frac{2 \pi i }{q} \frac{\delta_{i_y,j_y}}{\ell_y}.
		\label{32}
	\end{equation}
	
	The boundary action can be constructed by demanding that it gives rise to the above commutations. Thus, it must include the terms
	\begin{equation}
		S_{\text{boundary}} = \int dt\, \ell_y\sum_{i_y\,\in\,\text{odd}} \frac{q}{2 \pi} \left[  \partial_t\varphi_{L_x+1,i_y} \Delta_y\theta_{L_x,i_y+1} + \partial_t\theta_{L_x+1,i_y+1} \Delta_y\varphi_{L_x,i_y}+ \cdots \right].
		\label{boundaryaction}
	\end{equation}
	Additional terms follow from symmetry requirements. Symmetry operators \eqref{ert2} can be placed at the boundary, where they become
	\begin{eqnarray}
		&&\exp\left( i\, m^{y}\, \ell_y\displaystyle\sum_{i_y\,\in\,\text{odd}} \Delta_y\varphi_{L_x, i_y}\right),~~~ 	\exp\left( i\, m^{y}\, \ell_y\sum_{i_y\,\in\,\text{odd}} \Delta_y\varphi_{L_x+1, i_y}\right), \nonumber\\ 
		&& \exp\left( i\, n^{y}\, \ell_y\sum_{i_y\,\in\,\text{odd}} \Delta_y\theta_{L_x,i_y+1}\right),~~\exp\left( i\, n^{y}\, \ell_y\sum_{i_y\,\in\,\text{odd}} \Delta_y\theta_{L_x+1,i_y+1}\right).
	\end{eqnarray}
	These operators act nontrivially on the vertex operators in \eqref{vertex}. For example, 
	\begin{eqnarray}
		\exp\left( i\, n^{y}\, \ell_y\sum_{i_y\,\in\,\text{odd}} \Delta_y\theta_{L_x+1,i_y+1}\right)&& \exp \left(i \varphi_{L_x,i_y}\right)  \exp\left(- i\, n^{y}\, \ell_y\sum_{i_y\,\in\,\text{odd}} \Delta_y\theta_{L_x+1,i_y+1}\right)\nonumber\\
		= &&\exp i \left(\varphi_{L_x,i_y}+\frac{2 \pi }{q}n^y\right).
	\end{eqnarray}
	This amounts to a  $\mathbb{Z}_q$-shift in the field, i.e., 
	\begin{equation}
		\varphi_{L_x,i_y} \rightarrow \varphi_{L_x,i_y}+\frac{2 \pi }{q}n^y.
	\end{equation}
	Identical $\mathbb{Z}_q$-shift transformations follow for all the boundary fields. So all the terms which are invariant under these symmetries are allowed in the boundary action \eqref{boundaryaction}.
	
	Spatial fluctuations of the fields at the boundary are taken into account by terms such as 
	\begin{equation}
		\left(\Delta_y\varphi_{L_x,i_y}\right)^2,~~~ \left(\Delta_y\varphi_{L_x+1,i_y}\right)^2,~~~ \left(\Delta_y\theta_{L_x,i_y+1}\right)^2,~~~  \left(\Delta_y\theta_{L_x+1,i_y+1}\right)^2.
	\end{equation}

	Gapping terms compatible with the shift symmetries involve Hermitian combinations (cosines) of the bosonic vertex operators $e^{i q \varphi_{L_x,i_y}}$,  $e^{-i q \varphi_{L_x+1,i_y}}$, $e^{-i q \theta_{L_x,i_y+1}}$, $e^{i q \theta_{L_x+1,i_y+1}}$. In addition, to produce stable gapped phases the action must include only simultaneously commuting cosine operators. 
	
	An important point to have in mind is that the lattice boundary Hamiltonians exhibit explicitly only the symmetries $	\Gamma_{L_x}^{\text{even}}(0,1)$ and $ \Gamma_{L_x}^{\text{odd}}(0,1)$, but not their Kramers-Wannier dual counterpart \cite{Ji:2019jhk}. On the other hand, a field theory description involving dual fields $\varphi$ and $\theta$ exhibits explicitly both the symmetries and their dual counterparts. Therefore, to compare with the lattice Hamiltonian we need to focus only on the symmetries of one of the fields, say $\varphi$. Thus, we identify the lattice symmetries generated by 
	\begin{equation}
		\Gamma_{L_x}^{\text{even}}(0,1)~~~\text{and}~~~ \Gamma_{L_x}^{\text{odd}}(0,1),
	\end{equation}
	with the shifts of the fields 
	\begin{equation}
		\varphi_{L_x,i_y} \rightarrow \varphi_{L_x,i_y}+\frac{2 \pi }{q}n^y~~~\text{and}~~~\varphi_{L_x+1,i_y} \rightarrow \varphi_{L_x+1,i_y}+\frac{2 \pi }{q}{n'}^{y},
		\label{mlp}
	\end{equation}
	for the case $q=2$. With this in mind, we discuss below all possible gapping terms and their matching with the phases discussed in \ref{Vertical Boundary}.

	Let us begin with a gapping term corresponding to the phase described by the Hamiltonian \eqref{eo1}, which contains two exact symmetries. Such phase is stabilized by the gapping terms,
	\begin{equation}
		S_1=  \int dt\, \ell_y\sum_{i_y\,\in\,\text{odd}}\lambda_1 \left[\cos(q\theta_{L_x,i_y+1})+\cos(q\theta_{L_x+1,i_y+1}) \right],
	\end{equation}
	so that the shift symmetries of $\varphi_{L_x,i_y}$ and $\varphi_{L_x+1,i_y}$ are exact in the ground state. The dual symmetries, i.e., the shifts of $\theta_{L_x,i_y+1}$ and $\theta_{L_x+1,j_y+1}$, are spontaneously broken since when $\lambda_1$  flows to the IR, the cosine operators pin the fields at one of their minima and consequently the ground state configuration is degenerate.
	
	The next phase, following the sequence of Sec. \ref{Vertical Boundary}, is the cluster model Hamiltonian \eqref{eo2a}. This SPT phase is described by the following gapping terms 
	\begin{equation}
		S_2=  \int dt\, \ell_y\sum_{i_y\,\in\,\text{odd}} \lambda_2  \left[\cos(q(\varphi_{L_x,i_y}+\theta_{L_x,i_y-1})) +\cos(q(\varphi_{L_x+1,i_y-2}+\theta_{L_x+1,i_y+1}))\right].
	\end{equation}
	That these operators give rise to a SPT phase can be understood from the fact that this phase supports projective representation of boundary symmetries. To see this, we consider initially the system with periodic boundary conditions and pick the term of the action with $i_y=1$,
	\begin{equation}
		\cos(q(\varphi_{L_x,1}-\theta_{L_x,L_y})) +\cos(q(\varphi_{L_x+1,L_y-1}+\theta_{L_x+1,2})).
	\end{equation}
	Now, if the system has a boundary along the $y$-direction, then locality of the interactions prevents such terms to enter the action. Accordingly, the associated vertex operators 
	\begin{equation}
		e^{i \varphi_{L_x,1}},~~~e^{i\theta_{L_x+1,2}},~~~e^{-i\varphi_{L_x+1,L_y-1}},~~~e^{-i \theta_{L_x,L_y}},
	\end{equation}
	will commute with the Hamiltonian and then are symmetries. However, as they do not commute among themselves, such symmetries are projectively realized at the boundary of the 1+1 dimensional system.

	The phases described by \eqref{234a} and \eqref{234a1} correspond to
	\begin{equation}
		S_3=  \int dt\, \ell_y\sum_{i_y\,\in\,\text{odd}} \lambda_3  \left[\cos(q\varphi_{L_x,i_y})+\cos(q\theta_{L_x,i_y+1})\right]
	\end{equation}
	and
	\begin{equation}
		S_4=  \int dt\, \ell_y\sum_{i_y\,\in\,\text{odd}} \lambda_4\left[\cos(q\varphi_{L_x+1,i_y})+\cos(q\theta_{L_x+1,i_y+1})\right]. 
	\end{equation}
	In each one of these actions, one of the shifts symmetries \eqref{mlp} is exact while the other one is spontaneously broken.
	
	The diagonal symmetry $\Gamma_{L_x}^{\text{even}}(0,1)\Gamma_{L_x}^{\text{odd}}(0,1)$ is exact in the boundary phase described by the lattice Hamiltonian \eqref{111a}. It corresponds to the following gapping terms
	\begin{equation}
		S_5=  \int dt\, \ell_y\sum_{i_y\,\in\,\text{odd}} \lambda_5 \left[\cos(q(\varphi_{L_x,i_y}-\varphi_{L_x+1,i_y}))+\cos(q(\theta_{L_x,i_y+1}+\theta_{L_x+1,i_y+1}))\right].
	\end{equation}
	We see that the diagonal symmetry, which corresponds to the shifts in \eqref{mlp} with $n^y={n'}^y$, is exact.

	Finally, the full spontaneous symmetry breaking phase arising from the Hamiltonian \eqref{tyu} corresponds to 
	\begin{equation}
		S_6=  \int dt\, \ell_y\sum_{i_y\,\in\,\text{odd}} \lambda_6  \left[\cos(q\varphi_{L_x,i_y})+\cos(q\varphi_{L_x+1,i_y})\right].
	\end{equation}
	Both shift symmetries \eqref{mlp} are spontaneously broken.


	\section{Effective Field Theory: Low Energies and Long Distances}\label{eft02}
	
	We proceed to construct a low-energy and long-distance effective field theory capturing universal properties of the topological phase. The ideia is that in the continuum all the restrictions on the mobility of the excitations disappear, so that an effective theory can be constructed in terms of a set of gauge fields coupled by a $K$-matrix, which is chosen to encode the anyonic mutual statistics of the anyons. The ground state dependence on the lattice details enter the continuum theory through twisted boundary conditions of the  gauge fields on the torus of sizes
	\begin{equation}
		\mathscr{L}_x\equiv \ell_x L_x ~~~ \text{and} ~~~\mathscr{L}_y \equiv \ell_y L_y.
		\label{lengths}
	\end{equation}

	Since the lattice model contains four independent anyons, $e_1$, $e_2$, $o_1$, and $o_2$, we consider four gauge fields to represent them $a^I$, $I=1,\ldots,4$. The mutual Chern-Simons action is 
	\begin{equation}
		S = \int \,  \frac{K_{IJ}}{4 \pi} a^I da^J,
	\end{equation}
	with the $K$-matrix chosen as
	\begin{equation}
		K=\begin{pmatrix}~~
			2\, \sigma_x  & 0_{2\times 2}  \\
			0_{2\times 2} &  ~2\, \sigma_x ~~
		\end{pmatrix},
		\label{kmatrix}
	\end{equation}
	where $\sigma_x$ is a $x$-Pauli matrix acting on the space of fields. This $K$-matrix implies $\mathbb{Z}_2$ mutual statistics between $\{a^1,a^2\}$ and $\{a^3,a^4\}$. In this way, we associate 
	\begin{equation}
		a^1 \leftrightarrow e_1,~~~
		a^2 \leftrightarrow o_2,~~~
		a^3 \leftrightarrow o_1, ~~~
		a^4 \leftrightarrow e_2,
		\label{ident}
	\end{equation}
	reproducing the lattice mutual statistics.  
	
	To understand the ground state degeneracy, we shall analyze three distinct cases.

	
	\subsubsection{Nontwisted Boundary Conditions}

	We have seen that when $L_x$ is a multiple of three and $L_y$ is even, the anyons return to their initial position when transported around any one of the directions. From the continuum, this amounts to impose nontwisted boundary conditions in both directions
	\begin{equation}
		a^{I}(x+\mathscr{L}_x,y)=a^I(x,y)~~~\text{and}~~~ a^{I}(x,y+\mathscr{L}_y)=a^I(x,y).
		\label{rrr}
	\end{equation}
	This corresponds to the usual case of periodic boundary conditions and the ground state degeneracy is simply $\det K=16$.

	
	\subsubsection{Vertical Twisted Boundary Conditions}

	When  $L_x$ is a multiple of three and $L_y$ is odd, anyons must be transported twice around the $y$-direction to return to their initial positions. This characteristic can be incorporated in the continuum theory by choosing the boundary conditions along the $x$-direction as in \eqref{rrr}, but twisted in the $y$-direction, namely, 
	\begin{equation}
		a^{1}(x,y+\mathscr{L}_y)=a^3(x,y)~~~\text{and}~~~ a^{3}(x,y+\mathscr{L}_y)=a^1(x,y)
		\label{hhh}
	\end{equation}
	and
	\begin{equation}
		a^{2}(x,y+\mathscr{L}_y)=a^4(x,y)~~~\text{and}~~~a^{4}(x,y+\mathscr{L}_y)=a^2(x,y).
		\label{hhh1}
	\end{equation}
	These conditions imply 
	\begin{equation}
		a^I(x,y+2\mathscr{L}_y)=a^I(x,y).
	\end{equation}
	Consequently, the Fourier mode expansion of the fields reads
	\begin{equation}
		a_{\mu}^I(t,x,y)= \sum_{\vec{k}} e^{i \vec{k}\cdot\vec{x}}  \tilde{a}_{\mu}^I(t,k_x,k_y),
	\end{equation}
	with $k_x = \frac{2\pi}{\mathscr{L}_x} n_x $ and $k_y = \frac{\pi}{\mathscr{L}_y} n_y $, and $n_x, n_y\,\in\,\mathbb{Z}$. The twisted boundary conditions \eqref{hhh} and \eqref{hhh1} yield to
	\begin{equation}
		e^{i k_y \mathscr{L}_y}\tilde{a}_{\mu}^1(t,k_x,k_y)= \tilde{a}_{\mu}^3(t,k_x,k_y) ~~~\text{and}~~~ e^{i k_y \mathscr{L}_y}\tilde{a}_{\mu}^2(t,k_x,k_y)= \tilde{a}_{\mu}^4(t,k_x,k_y).
	\end{equation} 
	In particular, these relations imply
	\begin{equation}
		\tilde{a}_{\mu}^1(t,0)=\tilde{a}_{\mu}^3(t,0)~~~\text{and}~~~\tilde{a}_{\mu}^2(t,0)=\tilde{a}_{\mu}^4(t,0).
		\label{zeromodes}
	\end{equation}
	Thus, only two zero modes, say $\tilde{a}_{\mu}^1(t,0)$ and $\tilde{a}_{\mu}^2(t,0)$, are independent, and satisfy
	\begin{equation}
		[\tilde{a}_x^1, \tilde{a}_y^2]=\frac{\pi i }{2\mathscr{L}_x \mathscr{L}_y }~~~\text{and}~~~ [\tilde{a}_y^1, \tilde{a}_x^2]=-\frac{\pi i }{2\mathscr{L}_x \mathscr{L}_y }.
		\label{czm}
	\end{equation}
	
	Now it is clear to see how the twisted boundary conditions are reflected in the ground state degeneracy. To produce proper closed lines (symmetry operators) along the $y$-direction, the lines need to go twice around the system,
	\begin{equation}
		\exp \left(i \int_0^{2\mathscr{L}_y} dy\, a_y^I \right),
	\end{equation}
	whereas in the $x$-direction they need to go only once
	\begin{equation}
		\exp \left(i \int_0^{\mathscr{L}_x} dy\, a_x^I \right).
	\end{equation}
	The ground state degeneracy is dictated from the algebra of line operators involving only the zero modes, as the nonzero ones are just gauge freedom. Thus, the above line operators reduce to
	\begin{equation}
		\exp \left(i \int_0^{2\mathscr{L}_y} dy\, a_y^I \right)~ \rightarrow ~  	e^{i 2\mathscr{L}_y  \tilde{a}_y^I }~~~\text{and}~~~
		\exp \left(i \int_0^{\mathscr{L}_x} dy\, a_x^I \right) ~ \rightarrow ~  	e^{i \mathscr{L}_x  \tilde{a}_x^I }. 
	\end{equation}
	Finally, using \eqref{czm}, we obtain 
	\begin{equation}
		e^{i 2\mathscr{L}_y  \tilde{a}_y^1 } \,	e^{i \mathscr{L}_x  \tilde{a}_x^2}=  - e^{i \mathscr{L}_x  \tilde{a}_x^2}\, e^{i 2\mathscr{L}_y  \tilde{a}_y^1 }~~~\text{and}~~~
		e^{i 2\mathscr{L}_y  \tilde{a}_y^2 } \,	e^{i \mathscr{L}_x  \tilde{a}_x^1}=  - e^{i \mathscr{L}_x  \tilde{a}_x^1}\, e^{i 2\mathscr{L}_y  \tilde{a}_y^2 },
	\end{equation}
	which leads to a fourfold ground state degeneracy.

	
	\subsubsection{Horizontal Twisted Boundary Conditions}

	When $L_x$ is not a multiple of three, as the anyons are transported around the $x$-direction, they are permuted as
	\begin{align}
		e_1 \, \, &\xrightarrow{L_x} \, \, e_2 \xrightarrow{L_x} \, \, e_1e_2 \xrightarrow{L_x} \, \, e_1 \nonumber\\
		o_1 \, \, &\xrightarrow{L_x} \, \, o_2 \xrightarrow{L_x} \, \, o_1o_2 \xrightarrow{L_x} \, \, o_1.
	\end{align}
	According to the identifications in \eqref{ident}, these permutations can be implemented in the continuum through the following twisted boundary conditions
	\begin{equation}
		a^1(x+\mathscr{L}_x)=a^4(x)~~~\text{and}~~~ a^4(x+\mathscr{L}_x)=a^1(x)-a^4(x)
		\label{ggg0}
	\end{equation}
	and
	\begin{equation}
		a^2(x+\mathscr{L}_x)=a^2(x)+a^3(x) ~~~\text{and}~~~ 	a^3(x+\mathscr{L}_x)=a^2(x),
		\label{ggg1}
	\end{equation}
	where we are omitting the $y$-dependence. The minus sign in the second condition of \eqref{ggg0} is important to properly ensure the invariance of the action under spatial translations.  
	
	With these twisted boundary conditions, it is simple to see that all lines along the $x$-direction trivialize. For example, 
	\begin{eqnarray}
		\exp\left( i \int_{0}^{3\mathscr{L}_x} dx\, a^1_x \right)&=& 
		\exp\left( i \int_{0}^{\mathscr{L}_x} dx\, a^1_x(x)+i \int_{\mathscr{L}_x}^{2\mathscr{L}_x} dx\, a^1_x(x)+i \int_{2\mathscr{L}_x}^{3\mathscr{L}_x} dx\, a^1_x(x) \right)\nonumber\\
		&=&  \exp\left( i \int_{0}^{\mathscr{L}_x} dx\, \big{(}a^1_x(x)+a^1(x+\mathscr{L}_x)+a^1(x+2\mathscr{L}_x)\big{)} \right)\nonumber\\ 
		&=& \exp\left(  2i \int_{0}^{\mathscr{L}_x} dx\, a^1_x(x) \right),
	\end{eqnarray}
	where in the last step we have used \eqref{ggg0} recursively. Therefore, we see that this line is identified with the identity, since it has trivial commutation with all other lines. The same goes for all other lines along the $x$-direction. Consequently, there is no nontrivial algebra between line operators and the ground state is unique. 
	
	To close this section, we note that twisted boundary conditions can be interpreted in terms of defects. For example, a twist as we go along the $y$-direction can be thought as due to the presence of cut line extended in the perpendicular direction. The ground state degeneracy, in its turn, is the number of inequivalent anyons under the twist. When there is no defect, the number of inequivalent anyons is 16. For a twist as we go along the vertical direction, the number of inequivalent anyons is four and consequently the ground state if fourfold degenerate. Under a twist as we go along the $x$-direction all the anyons are equivalent and the ground state is unique.


	\section{$h$-Partite Generalized Modulated Symmetries}\label{hpartitecase}
	
	The tripartite model can be generalized in order to describe a large class of $\mathbb{Z}_2$ topological ordered phases. In this context, we can use straightly most of the methods discussed previously, so that we shall be brief in this section. 
	
	To proceed, we recall \eqref{eq:const2}, which corresponds to a recurrence relation along the $x$-direction following from the constraints between the commuting projectors. It can be generalized to a linear recurrence relation of odd size $h$: 
	\begin{equation}
		t_{\vec{r}}+t_{\vec{r}+\hat{x}}+ \cdots +t_{\vec{r}+(h-1)\hat{x}} = 0 \mod 2
		\label{eq:ma1}
	\end{equation}
	and
	\begin{equation}
		t_{\vec{r}+\hat{y}}+t_{\vec{r}-\hat{y}} = 0 \mod 2.
	\end{equation}
	Assuming that $L_x$ is a multiple of $h$, the horizontal constraints lead to a set of $h-1$ linearly independent solutions, which can be chosen as
	\begin{equation}
		\label{8.3}
		\underbrace{(1,1,0,0,0,\ldots,0)}_{h\,\, \text{sites}}, (1,0,1,0,0\ldots,0), (1,0,0,1,0,\ldots,0),\ldots,(1,0,0,0,\ldots,1).
	\end{equation}
	It is simple to verify that this set generates any other solution of \eqref{eq:ma1}.

	The lattice Hamiltonian that reproduces the recurrence relation \eqref{eq:ma1} is
	\begin{equation}
		H = - \sum_{\vec{r}} P^{(h)}_{\vec{r}},
	\end{equation}
	where $P_{\vec{r}}^{(h)}$ are commuting projectors defined by
	\begin{equation}
		P^{(h)}_{\vec{r}} = \underbrace{ \begin{matrix}
				{}&{}&{} & X &{} &{}&{}
				\\
				\cdots&Z & Z & Z_{\vec{r}} & Z &Z& \cdots
				\\
				{}&{}&{} & X &{} &{} &{}
		\end{matrix}}_{h\, \text{consecutive sites}}.
	\end{equation}
	
	By following the discussions of Sec. \ref{tripartite}, the ground state degeneracy can be computed in a simple way. For the case of $L_x$ being a multiple of $h$, it involves the $h-1$ linearly independent solutions of the constraints along the $x$-direction,
	\begin{equation}
		\text{GSD} = 2^{\gcd(2,L_y)(h-1)}.
		\label{simplecase}
	\end{equation}
	When $L_x$ is not a multiple of $h$, we need to take into account that it may happen that $\gcd(h,L_x)\neq 1$, which leads to nontrivial solutions of the constraints \eqref{eq:ma1}. In the general case, the ground state degeneracy is
	\begin{equation}
		\text{GSD} = 2^{\gcd(2,L_y)\left[\gcd(h,L_x)-1\right]}.
		\label{generalcase}
	\end{equation}
	Note that for any $h$, there are always sizes which have a nondegenerate ground state, namely, when $\gcd(h,L_x) = 1$.
	
	
	\subsection{Symmetry Operators}
	
	Most of the results for a general size $h$ follows from the straight generalization of the tripartite case. For simplicity, we shall consider the case where $L_x$ is a multiple of $h$ and $L_y$ is even. 
	
	The horizontal symmetry operators are products of $X$-operators disposed according to the solutions of the recurrence relation, and they are labeled by their odd and even $i_y$ position,
	\begin{equation}
		W_{i_y \, \in \, \text{even}}^{(h)} = \prod_{i_x=1}^{L_x}X_{(i_x,i_y)}^{t_{(i_x,i_y)}}~~~\text{and}~~~ W_{i_y \,\in\, \text{odd}}^{(h)} = \prod_{i_x=1}^{L_x}X_{(i_x,i_y)}^{t_{(i_x,i_y)}}.
	\end{equation}
	
	Following the notation of Sec. \ref{symmetryperators}, we can distinguish these operators by the occupation numbers $(n_{i_x-(h-1)},n_{i_x-(h-2)}, \ldots, n_{i_x})$, which determine whether a given site is acted by an $X$-operator\footnote{For the case $h=3$ studied in  Sec. \ref{symmetryperators}, we have considered only two occupation numbers to specify the extended operator, but in the present discussion we will keep  an additional redundant number to make explicit the relation with (\ref{8.3}).}. In this way, we specify the above operators as
	\begin{equation}
		W_{i_y \, \in\, \text{even}}(n_{i_x-(h-1)},n_{i_x-(h-2)}, \ldots, n_{i_x}) \quad \text{and} \quad 	W_{i_y \,\in\, \text{odd}}(n_{i_x-(h-1)},n_{i_x-(h-2)}, \ldots, n_{i_x}).
	\end{equation}
	Note that there are $h-1$ independent operators for each one of the sublattices. When the size $L_y$ is odd, there is no distinction between these two symmetries, as they can be deformed into each other. If $L_x$ is not a multiple of $h$, the number of independent symmetry lines reduces to $\gcd(L_x,h)-1$.
	
	Vertical symmetry operators are alternating products of $Z$'s, 
	\begin{equation}
		\Gamma_{i_x}^{\text{even}} = \prod_{i_y \,\in\, \text{even}} Z_{(i_x,i_y)} \quad \text{and} \quad \Gamma_{i_x}^{\text{odd}} = \prod_{i_y \,\in\, \text{odd}} Z_{(i_x,i_y)}.
	\end{equation}
	They can also be specified in terms of  occupation numbers $(m_{i_x-(h-1)},m_{i_x-(h-2)}, \ldots, m_{i_x})$, which determine whether a vertical line is crossing a given 
	$x$-position. In this way, we represent they as 
	\begin{equation}
		\Gamma_{i_x}^{\text{even}}(m_{i_x-(h-1)},m_{i_x-(h-2)}, \ldots, m_{i_x}) \quad \text{and} \quad \Gamma_{i_x}^{\text{odd}}(m_{i_x-(h-1)},m_{i_x-(h-2)}, \ldots, m_{i_x}).
	\end{equation}

	Open horizontal and vertical operators can be used to transport excitations. The vertical transport is independent of the support $h$ of the commuting projectors and is restricted to the even and odd sublattices for $L_y$ even, as discussed in Sec. \ref{symmetryperators}. 
	
\begin{figure}
		\centering
		\includegraphics[angle=90,scale=0.65]{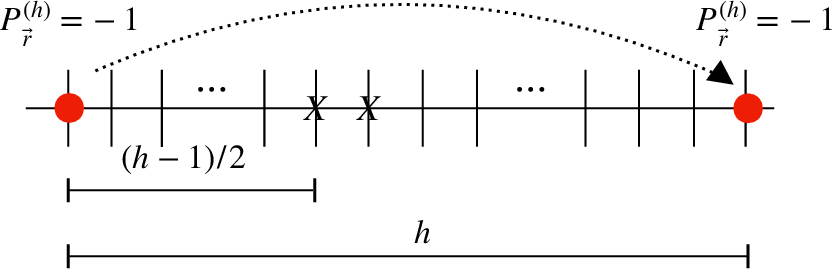}
		\caption{The red balls indicate the initial and final positions of an excitation in a transport process. The picture shows that an excitation can move along the $x$-direction only in rigid steps of size $h$ in units of the lattice spacing. }
		\label{transporth}
	\end{figure}

	On the other hand, the horizontal transport, carried by open operators of the form $W_{i_y \,\in\, \text{even}}(1,1,0, \ldots, 0)$ and $W_{i_y \,\in\, \text{odd}}(1,1,0, \ldots, 0)$, takes place in rigid steps of size $h$, as illustrated in Fig. \ref{transporth}. In a region of size $h$ units of the lattice spacing, we can have $h$ excitations but only $h-1$ of them are independent (in the sense that given $h-1$ excitations, the last one can be obtained through the application of a local operator - they belong to the same superselection sector. As the $h-1$ excitations cannot be connected by transport, they actually correspond to distinct excitations, characterized by the position. When $L_x$ is not a multiple of $h$, with $\gcd(L_x,h)=1$, it turns out that all the excitations are in the same superselection sector, i.e., there is a single superselection sector, as there are extended operators creating local excitations in each of the positions.

	To conclude this section, it is opportune to mention that the topological entanglement entropy can be computed in a similar way as in Sec. \ref{entanglement}, resulting in 
	\begin{equation}
		S_{\text{topo}}=(h-1)\log 2.
	\end{equation}
	According to the relation \eqref{990099}, this result matches with the particle spectrum of $2^{2(h-1)}$ anyons, following from all possible combinations of the maximum number $2(h-1)$ of basic excitations ($h-1$ in the horizontal even and $h-1$ in the horizontal odd sites).

	\subsection{'t Hooft Anomalies}
	
	As discussed in \ref{thooftsection}, the ground state degeneracy is a reflection of 't Hooft anomalies. For example, for the case in which $L_x$ is a multiple of $h$ and $L_y$ is even, the computation of the 't Hooft anomalies is immediate. We simply choose a basis for the occupation numbers of $\Gamma_{i_x}^{\text{even}}(n_{i_x-(h-1)},n_{i_x-(h-2)}, \cdots, n_{i_x})$ as
	\begin{equation}
		\{(0,1,0, \cdots, 0), (0,0,1, \cdots, 0), \cdots, (0,0,0, \cdots, 1)\},
	\end{equation}
	and for $W_{i_y \,\in\, \text{even}}(n_{i_x-(h-1)},n_{i_x-(h-2)}, \cdots, n_{i_x})$, as 
	\begin{equation}
		\{(1,1,0, \cdots, 0), (1,0,1, \cdots, 0), \cdots, (1,0,0, \cdots, 1)\}.
	\end{equation}
	Now it is easy to identify the commutation relation between these symmetry operators. We just have to compare the occupation numbers of the two types of operators. Whenever 
	two operators $\Gamma_{i_x}^{\text{even}}$ and $W_{i_y \,\in\, \text{even}}$ share a nonzero occupation number at the same position, this implies that they anti-commute,  
	\begin{equation}
		\big\{W_{i_y \,\in\, \text{even}}(\underbrace{1,0, \cdots, 0}_{j  \, \text{times}},1,0, \cdots), \, \Gamma_{i_x}^{\text{even}}(\underbrace{0,0, \cdots, 0}_{j  \, \text{times}},1,0, \cdots)\big\} = 0.
	\end{equation}                           
	For the even sublattice, this gives rise to $h-1$ anomalies. Taking into account the odd sublattice, we get a total of $2(h-1)$ anomalies, which lead to the ground state degeneracy \eqref{simplecase}.  This discussion can be straightforwardly extended to recover the general case  \eqref{generalcase}.

	\section{Final Remarks}\label{finalremarks}

	In this work we have discussed classes of $\mathbb{Z}_2$ topologically ordered phases characterized by generalized modulated symmetries. 
	Explicit realizations of such phases are provided by fixed-point Hamiltonians involving commuting projectors with support in a region of size $h$ along the horizontal direction. The modulation of the symmetries are dictated by the support $h$ of the commuting projectors.

	The modulated symmetries are sensitive to the lattice sizes. For sizes with $\gcd(L_x,h)\neq 1$, they are spontaneously broken, leading to a nontrivial ground state degeneracy. This degeneracy can also be understood in terms of  t'Hooft anomalies between the symmetry operators, preventing a unique ground state. 
	On the other hand, when $\gcd(L_x,h)=1$, the modulated symmetries are explicitly broken by the lattice size, and there are no nontrivial symmetry operators (nor 't Hooft anomalies). Consequently, the ground state is unique. In this case, there are extended operators\footnote{The would-be symmetry operators, but which fail to properly close under periodic boundary conditions.} that create local excitations, causing all anyons to belong to the same superselection sector. Such interplay between the lattice size and the ground state degeneracy is a manifestation of UV/IR mixing.

	The structure of the underlying modulated symmetries implies that the anyons can only move in rigid steps of size $h$, which in turn leads to the existence of position-dependent $\mathbb{Z}_2$ anyons. This is in contrast with some previous studies with position-dependent anyons \cite{Oh_2022,Pace_2022,Delfino_2023,Watanabe:2022pgk,Oh:2023bnk}, arising in $\mathbb{Z}_N$ topologically ordered phases. In such cases, the characteristic step-size of the anyons is dictated by the order $N$ of the symmetry $\mathbb{Z}_N$.

	We explored extensively the simplest case $h=3$, which captures the essence of the involved physics. We carried a detailed study of the boundary physics, and show that it exhibits a rich variety of gapped phases, including trivial, partial and total symmetry-breaking, and SPT phases. In addition, we studied effective field theory descriptions, which provide enlightening perspectives into the problem. In this context, we first derived a low-energy and short-distance description, which makes transparent the relation between the generalized modulated symmetries and the restrictions on anyon mobility, and incorporates the boundary physics in a natural way. Then, we constructed a low-energy and long-distance description, that shows how the short-distance information can be codified in the continuum through the twisted boundary conditions.

	We conclude by noting that the phases discussed here combine the simplicity of fixed-point Hamiltonians with remarkable physical properties, so that they can serve a variety of purposes for further theoretical investigations.

	
	\begin{acknowledgments}

		PRSG would like to thank Guilherme Delfino for useful conversations. We also acknowledge the financial support from the Brazilian funding agencies CAPES and CNPq. 
		
	\end{acknowledgments}


	
	\appendix


	
	\section{Horizontal Boundary Phases}\label{appendixA}
	
	In this appendix, we construct the phases corresponding to the condensation of the set of anyons given in \eqref{ind} and \eqref{comp}, for the case of a  horizontal boundary. In this case, we consider $L_x$ being a multiple of three and the lattice has to be partitioned in three. The boundary operators are
	\begin{align}
		B^{h}_{1, i_x}=\,\begin{matrix}
			Z & Z_{i_x} & Z \\
			{} & X &{}	
		\end{matrix}
		\qquad\qquad\text{and} \qquad\qquad B_{2,i_x}^{h}=X_{i_x}. 
		\label{stab horizontal}
	\end{align}
	
	As in Sec. \ref{Vertical Boundary}, we look for Hamiltonians constructed from combinations of these operators, which lead to distinct gapped phases of condensation of the anyons.

	
	\subsubsection{Condensation of $\{ e_1,e_2 \}$}
	
	The first Hamiltonian we consider is a sum of $B^{h}_{2, i_x}$ over the boundary and is given by
	\begin{equation}
		H^h(e_1,e_2) = - \sum_{i_x} X_{i_x}.
	\end{equation}
	The ground state is a paramagnet and it gives rise to a trivially gapped phase. This Hamiltonian condenses the pair of anyons $\{ e_1,e_2 \}$, as we can see from the symmetry lines $\Gamma_{j_x}^{\text{odd}}(1,0)$ and $\Gamma_{j_x}^{\text{odd}}(0,1)$, where $j_x = 2 \bmod 3$,  associated with the transport of that pair of anyons, which commute with the Hamiltonian $H^h(e_1,e_2)$.
	
	
	\subsubsection{Condensation of $\{e_1o_1,e_2o_2\}$}
	
	The next Hamiltonian is built from a product of both $B^{h}_{1, i_x}$ and $B^{h}_{2, i_x}$,
	\begin{equation}
		H^h(e_1o_1,e_2o_2) =-\sum_{i_x} \,\begin{matrix}
			{}Z &Y_{i_x} &Z
			\\
			&X
		\end{matrix}.
	\end{equation}
	The line operators that correspond to the pair of anyons $\{e_1o_1,e_2o_2\}$ are $\Gamma_{j_x}^{\text{even}}(1,0)\Gamma_{j_x}^{\text{odd}}(1,0)$ and $\Gamma_{j_x}^{\text{even}}(0,1)\Gamma_{j_x}^{\text{odd}}(0,1)$. We can see that they commute with this boundary Hamiltonian leading to the condensation of $\{e_1o_1,e_2o_2\}$.
	
	The corresponding boundary phase can be analyzed by the effective Hamiltonian that freezes the spins located at $i_y=L_y-1$, as discussed in Sec. \ref{Vertical Boundary},
	\begin{equation}
		H_{\text{eff}}^h(e_1o_1,e_2o_2) =-\sum_{i_x} \begin{matrix}
			{}Z &Y_{i_x} &Z.
		\end{matrix}
	\end{equation}
	This is the Hamiltonian of the cluster model  \cite{2006RpMP...57..147N,Son_2011}, which is a SPT phase protected by the symmetries $W_{L_y}(1,0)$ and $W_{L_y}(0,1)$.
	
	
	\subsubsection{Condensation of $\{e_3,o_3\}$}
	
	The third Hamiltonian corresponds to the condensation of the pair $\{e_3,o_3\}$
	\begin{equation}
		H^h(e_3,o_3) = -\sum_{i_x =\, 0 \bmod 3} \left( \begin{matrix}
			{}Z &I_{i_x} &I &Z
			\\
			&X & X
		\end{matrix}
		\, \, + X_{i_x}+X_{i_x+1} \right).
	\end{equation}
	The symmetry operators related to the mobility of these anyons are $\Gamma_{j_x+1}^{\text{odd}}(0,1)$ and $\Gamma_{j_x+1}^{\text{even}}(0,1)$. These operators commute with $H^h(e_3,o_3)$, leading to the condensation of the anyons associated with them.
	
	We can see that the corresponding ground states realize the SSB of a $\mathbb{Z}_2$ symmetry. The effective Hamiltonian, obtained by integrating out the bulk degrees of freedom, reads
	\begin{equation}
		H_{\text{eff}}^h(e_3,o_3) = -\sum_{i_x = \,0 \bmod 3}\left( \begin{matrix}
			{}Z &I_{i_x} &I &Z
		\end{matrix}
		\, \, + X_{i_x}+X_{i_x+1}\right).
	\end{equation}
	This Hamiltonian has a local symmetry operator given by $Z_{i_x}$, for $i_x = 2 \bmod 3$. Note that it does not commute with the symmetry operators $W_{L_y}(1,0)$ and  $W_{L_y}(0,1)$ at the boundary, but it does commute with $W_{L_y}(1,1)$. So  the linear combination 
	\begin{equation}
		\Psi_1 \equiv \frac{3}{L_x} \sum_{i_x = 2 \bmod 3} Z_{i_x}
	\end{equation}
	is able to distinguish the ground states and has the following algebra 
	\begin{align}
		\{\Psi_1,W_{L_y}(1,0)&\}=0, \quad \{\Psi_1,W_{L_y}(0,1)\}=0,\\
		&[\Psi_1,W_{L_y}(1,1)] = 0.
	\end{align}
	
	To further understand this point, we can pick the two degenerate ground states as
	\begin{equation}
		\label{432}
		\bigotimes_{i_x=0 \bmod 3}
		\ket{ \begin{matrix}
				{}\uparrow_{i_x-1}, &+, &+ 
		\end{matrix}}
		\quad \text{and} \quad 
		\bigotimes_{i_x=0 \bmod 3}
		\ket{ \begin{matrix}
				{}\downarrow_{i_x-1}, &+, &+ 
		\end{matrix}}.
	\end{equation}
	The symmetry operators $W_{L_y}(1,0)$ and  $W_{L_y}(0,1)$ act nontrivially on these ground states, while $\Psi_1$ distinguishes them by their eigenvalues.
	
	
	\subsubsection{Condensation of $\{e_1,o_1\}$}
	
	The next Hamiltonian is a translation of the previous ones,
	\begin{equation}
		H^h(e_1,o_1) = -\sum_{i_x = 1 \bmod 3}\left( \begin{matrix}
			{}Z &I_{i_x} &I &Z
			\\
			&X & X
		\end{matrix}
		\, \, + X_{i_x}+X_{i_x+1}\right).
	\end{equation}
	It leads to the condensation of the pair $\{e_1,o_1\}$, which can be seen from fact that the symmetry operators $\Gamma_{j_x}^{\text{odd}}(1,0)$ and  $\Gamma_{j_x}^{\text{even}}(1,0)$ commute with the boundary Hamiltonian.
	
	The analysis for this Hamiltonian is very similar to the previous one. There is a local operator that commutes with $H^h(e_1,o_1)$, given by $Z_{i_x}$, with $i = 0 \bmod 3$. It does not commute with $W_{L_y}(1,0)$ and $W_{L_y}(0,1)$, but it commutes with $W_{L_y}(1,1)$. The linear combination
	\begin{equation}
		\Psi_2 \equiv \frac{3}{L_y} \sum_{i_x = 0 \bmod 3} Z_{i_x}
	\end{equation}
	allows to distinguish the ground states, which can be obtained from a translation of the ones in Eq. (\ref{432}), 
	\begin{equation}
		\bigotimes_{i_x=1 \bmod 3}
		\ket{ \begin{matrix}
				{}\uparrow_{i_x-1}, &+, &+ 
		\end{matrix}}
		\quad \text{and} \quad 
		\bigotimes_{i_x=1 \bmod 3}
		\ket{ \begin{matrix}
				{}\downarrow_{i_x-1}, &+, &+ 
		\end{matrix}}.
	\end{equation}
	These ground states are connected by the nontrivial action of $W_{L_y}(1,0)$ and $W_{L_y}(0,1)$, and are distinguished by $\Psi_2$.
	
	
	\subsubsection{Condensation of $\{e_2,o_2\}$}
	
	There is one more Hamiltonian obtained from a translation of $H^h(e_3,o_3)$, but this time leading to the condensation of the pair $\{e_2,o_2\}$
	\begin{equation}
		H^h(e_2,o_2) = -\sum_{i_x = 2 \bmod 3 } \left( \begin{matrix}
			{}Z &I_{i_x} &I &Z
			\\
			&X & X
		\end{matrix}
		\, \, + X_{i_x}+X_{i_x+1} \right).
	\end{equation}
	The analysis is very similar to the previous ones. There is local operator commuting with this Hamiltonian given by $Z_{i_x}$, with $i_x = 1 \bmod 3$, from which we define the linear combination
	\begin{equation}
		\Psi_3 \equiv \frac{3}{L_x} \sum_{i_x = 1 \bmod 3} Z_{i_x}.
	\end{equation}
	The algebra of this local operator with the other symmetry operators is
	\begin{align}
		\{\Psi_3,W_{L_y}(1,0)&\}=0, \quad \{\Psi_3,W_{L_y}(1,1)\}=0\\
		&[\Psi_3,W_{L_y}(0,1)] = 0.
	\end{align}
	Then, the ground states
	\begin{equation}
		\bigotimes_{i_x=2 \bmod 3}
		\ket{ \begin{matrix}
				{}\uparrow_{i_x-1}, &+, &+ 
		\end{matrix}}
		\quad \text{and} \quad 
		\bigotimes_{i_x=2 \bmod 3}
		\ket{ \begin{matrix}
				{}\downarrow_{i_x-1}, &+, &+ 
		\end{matrix}}
	\end{equation}
	break both $W_{L_y}(1,0)$ and $W_{L_y}(1,1)$, and are distinguished by $\Psi_3$.
	
	
	\subsubsection{Condensation of $\{o_1,o_2\}$}
	
	The last Hamiltonian we consider on the horizontal boundary condenses the pair $\{o_1,o_2\}$ and is given by
	\begin{equation}
		H^h(o_1,o_2) =-\sum_i \,\begin{matrix}
			{}Z &Z_{i_x} &Z
			\\
			&X
		\end{matrix}.
	\end{equation}
	
	The symmetry operators corresponding to the transport of these anyons to the boundary are $\Gamma_{j_x}^{\text{even}}(1,0)$ and $\Gamma_{j_x}^{\text{even}}(0,1)$ and they commute with $H^h(o_1,o_2)$. The effective Hamiltonian obtained by integrating out the bulk is
	\begin{equation}
		H_{\text{eff}}^h(o_1,o_2) =-\sum_i \begin{matrix}
			{}Z &Z_{i_x} &Z.
		\end{matrix}
	\end{equation} 
	
	This Hamiltonian has a local symmetry operator given by $Z_{i_x}$ for any point on the boundary, and it anti-commutes with every $W_{L_y}(n,m)$, in particular with $W_{L_y}(1,1)$ and $W_{L_y}(1,0)$. From this observation, we define two different linear combinations of the local symmetry operators
	\begin{equation}
		\Omega \equiv \frac{3}{L_x} \sum_{i_x = 2 \bmod 3} Z_{i_x}~~~\text{and}~~~  \Omega^{\prime}  \equiv \frac{3}{L_x} \sum_{i_x = 0 \bmod 3} Z_{i_x},
	\end{equation}
	satisfying the following algebra with the other symmetry operators
	\begin{align}
		&\{\Omega,W_{L_y}(1,1)\}=0, \quad [\Omega,W_{L_y}(1,0)]=0\\
		&[\Omega^{\prime},W_{L_y}(1,1)]=0, \quad \{\Omega^{\prime},W_{L_y}(1,0)\}=0.
	\end{align}
	
	So $\Omega$ and $\Omega^{\prime}$ allow to distinguish the four degenerate ground states expected from the $\mathbb{Z}_2\times\mathbb{Z}_2$ symmetry breaking. We can see this from the four ground states below
	\begin{align}
		&	\bigotimes_{i_x=1 \bmod 3}
		\ket{ \begin{matrix}
				{}\uparrow_{i_x}, &\uparrow, &\uparrow
		\end{matrix}}
		\quad \text{,} \quad 
		\bigotimes_{i_x=1 \bmod 3}
		\ket{ \begin{matrix}
				{}\uparrow_{i_x}, &\downarrow, &\downarrow 
		\end{matrix}} \\
		& \bigotimes_{i_x=1 \bmod 3}
		\ket{ \begin{matrix}
				{}\downarrow_{i_x}, &\downarrow, &\uparrow
		\end{matrix}}
		\quad \text{,} \quad 
		\bigotimes_{i_x=1 \bmod 3}
		\ket{ \begin{matrix}
				{}\downarrow_{i_x}, &\uparrow, &\downarrow
		\end{matrix}}.
	\end{align} 
	All these states can  be connected by the applications of $W_{L_y}(1,1)$ and $W_{L_y}(1,0)$, and they are distinguished by the pair of the eigenvalues of $\Omega$ and $\Omega^{\prime}$. 
	
	
	\section{Topological Entropy for Non-Contractible Boundaries}\label{appendixB}
	
	It has been shown in Sec. \ref{entanglement} that the topological entropy of the model with $h=3$ is non-zero (independent of the lattice size), signaling topological ordering. We discuss here the computation of the topological entanglement entropy for a non-contractible region on a torus, which will lead to a lattice size dependence. As explained in Ref. \cite{Kim:2023qqi}, the topological entropy for this type of region is captured by 
	\begin{equation}
		\label{B5}
		S_{\text{topo}} = 2 \log \frac{\mathcal{D}}{d_{a}},
	\end{equation}
	where $d_{a}$ is the quantum dimension of the defect $a$.  If there is no such a defect present in the system, $d_{a}=1$. This entropy is related to the number of inequivalent anyons under the defect. In particular, if there is no defect, the entropy is simply $\log \mathcal{D}^2$, which is the number of quasiparticles in an Abelian topological order.
	
\begin{figure}[h]
		\centering
		\includegraphics[width=0.5\linewidth]{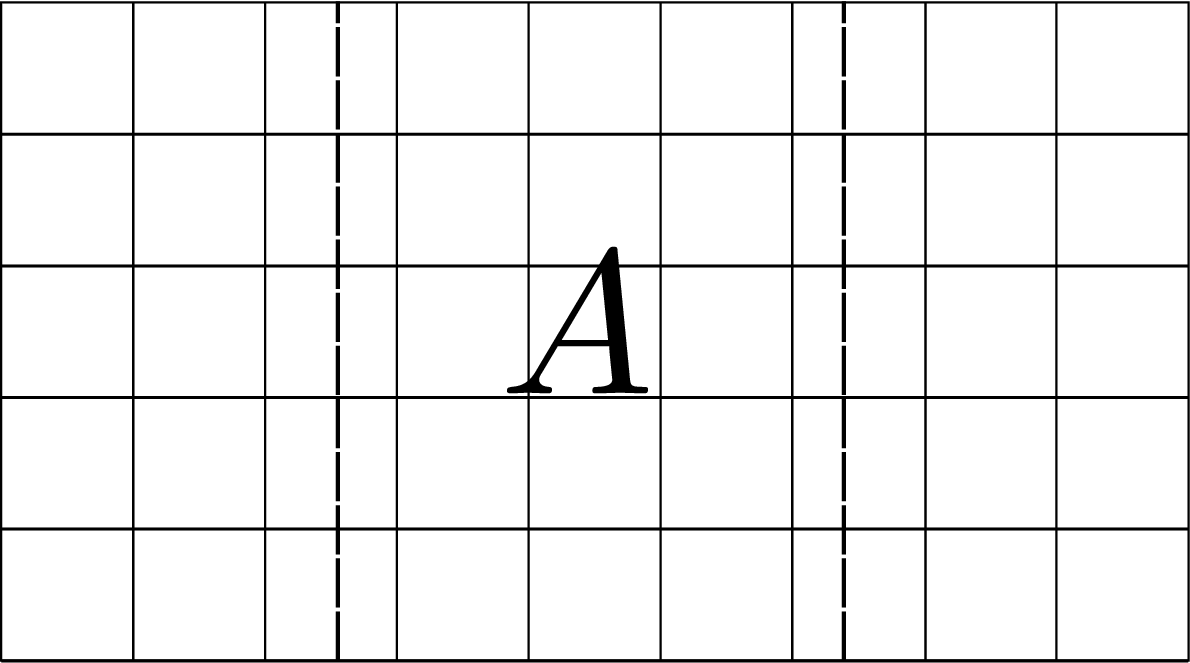}
		\caption{Non-contractible subregion $A$ that winds around the torus on the $y$-direction.}
		\label{fig:noncont}
	\end{figure}

	To proceed, we will calculate the entropy $S_A$ for the sub-region $A$ that winds around the torus in the $y$-direction, as shown in Fig. \ref{fig:noncont}. Then we shall extract the topological contribution, discarding perimeter contributions. 
	
	All we have to do is calculate the order of the group $G_A$ and use the result given in \eqref{ee01}. Note that each spin inside the region is in one-to-one correspondence with a stabilizer completely contained in $A$, except for those at the boundaries. If the horizontal size of the region $A$ is $l_x$, then the order of the group $G_A$ is 
	\begin{equation}
		\vert G_A \vert = 2^{l_x L_y-2L_y}.
	\end{equation}
	This result is not quite correct. This is because we are on a torus, and hence we can use operators outside the region $A$ to produce an operator contained in $A$. This gives additional operators to those counted in the above expression. If $L_y$ is even, we can take the product of the stabilizers on the even sublattice of the region outside $A$ in two different ways to produce an operator contained in $A$ (such products involve necessarily stabilizers on the boundary of the region $A$ and there are two ways a stabilizer can intercept the boundary; each way gives rise to a distinct operator). The same can be done for the odd sublattice. If $L_y$ is odd, the product has to be taken between operators in both even and odd lattices. Therefore, the correct order of $G_A$ is 
	\begin{equation}
		\vert G_A \vert = 2^{l_x L_y-2L_y}\times 2^{2\gcd{(L_y,2)}}.
	\end{equation}
	According to \eqref{ee01}, the entropy for the sub-region $A$ is
	\begin{equation}
		S_A = (2L_y-2\gcd{(L_y,2)})\log 2,
	\end{equation}
	and the topological contribution is given by
	\begin{equation}
		\label{B3}
		S_{\text{topo}}=\log 2^{2\gcd{(L_y,2)}}.
	\end{equation}
	This result matches the defect picture provided by expression \eqref{B5}. When the size $L_y$ is even, there is no twist acted upon the anyons when they traverse around the torus, but when the size is odd, the anyons are acted upon by exchanging $e_i$ with $o_i$. This result shows that the topological entropy for non-contractible loops is sensitive to the lattice size, which is a consequence of the UV/IR mixing.

	We can proceed similarly for a horizontal non-contractible region. As in the vertical case, we pick a rectangle of size $l_y$ and $L_x$. The naive order of group $G_A$ is
	\begin{equation}
		\vert G_A \vert = 2^{l_y L_x-2L_x}.
	\end{equation}
	Once again, this expression is missing some operators formed from stabilizers outside $A$. We can take the product between operators outside $A$ and at the boundaries. When the size $L_x$ is multiple of three, there are four independent ways to combine products between stabilizers to produce such operator (two combinations from products along the $x$-direction and two combinations corresponding to the two ways a stabilizer involved in the product can intercept the boundary). When $L_x$ is not a multiple of three, there is no way to produce an operator completely contained in $A$ from operators outside $A$. Putting all together, we obtain that the order of $G_A$ is
	\begin{equation}
		\vert G_A \vert = 2^{l_y L_x-2L_x}\times 2^{2(\gcd(L_x,3)-1)},
	\end{equation}
	and the topological entropy is given by
	\begin{equation}
		S_{\text{topo}} = \log{2^{2(\gcd{(L_x,3)}-1)}}.
	\end{equation}
	This result also matches the defect picture given in \eqref{B5}.

	\bibliography{references}
	
\end{document}